\documentclass[11pt]{amsart}
\usepackage[colorlinks=true,linkcolor=blue]{hyperref}
\usepackage{subfigure,caption2}
\usepackage{amssymb}
\usepackage{graphicx}
\usepackage[rflt]{floatflt}
\numberwithin{equation}{section}
\newcommand{\smallpagebreak}{{\par\vspace{2 mm}\noindent}}
\newcommand{\medpagebreak}{{\par\vspace{4 mm}\noindent}}

\newcommand{\demo}{\par\noindent{\it Proof.\/} \ }

\newcommand{\dsize}{\textstyle}
\newcommand{\D}{\displaystyle}

\newcommand{\R}{{\mathbb R}}

\newcommand{\Z}{{\mathbb Z}}
\newcommand{\N}{{\mathbb N}}
\newcommand{\C}{{\mathbb C}}
\newcommand{\Q}{{\mathbb Q}}

\newcommand{\re}{{\rm Re}\,}
\newcommand{\im}{{\rm Im}\,}

\newcommand{\res}{{\rm res}\, }
\newcommand{\Const}{{\rm Const}}
\newcommand{\Wc}{{\mathcal W}}

\theoremstyle{plain}
\newtheorem{Th}{Theorem}[section]
\newtheorem{Le}{Lemma}[section]
\newtheorem{Pro}{Proposition}[section]
\newtheorem{Cor}{Corollary}[section]
\theoremstyle{definition}
\newtheorem{Rem}{Remark}[section]
\newtheorem{Def}{Definition}[section]
\title{On the singular spectrum for adiabatic quasi-periodic
  Schr{\"o}dinger operators on the real line}
\author{Alexander Fedotov} \author{Fr{\'e}d{\'e}ric Klopp}
\address[Alexander Fedotov]{Department of Mathematical Physics, St
  Petersburg State University, 1, Ulia\-novskaja, 198904 St
  Petersburg-Petrodvorets, Russia}
\email{\href{mailto:fedotov@mph.phys.spbu.ru}{fedotov@mph.phys.spbu.ru}}
\address[Fr{\'e}d{\'e}ric Klopp]{D{\'e}partement de Math{\'e}matique, Institut
  Galil{\'e}e, U.R.A 7539 C.N.R.S, Universit{\'e} de Paris-Nord, Avenue J.-B.
  Cl{\'e}ment, F-93430 Villetaneuse, France}
\email{\href{mailto:klopp@math.univ-paris13.fr}{klopp@math.univ-paris13.fr}}
\keywords{quasi periodic Schr{\"o}dinger equation, Lyapunov exponent,
  singular spectrum, complex WKB method, monodromy matrix}
\subjclass{34E05, 34E20, 34L05}
\thanks{A.F. thanks the Universit{\"a}t Potsdam where part of this work
  was done. F.K.'s research was partially supported by the program
  RIAC 160 at Universit{\'e} Paris 13 and by the FNS 2000 ``Programme
  Jeunes Chercheurs''. Both authors thank the Mittag-Leffler Institute
  where part of this work was done.}
\begin{document}
\begin{abstract}
  In this paper, we study spectral properties of a family of
  quasi-periodic Schr{\"o}din\-ger operators on the real line in the
  adiabatic limit.  We assume that the adiabatic iso-energetic curves
  are extended along the momentum direction. In the energy intervals
  where this happens, we obtain an asymptotic formula for the Lyapunov
  exponent, and show that the spectrum is purely singular.
  \vskip.5cm
  \par\noindent   \textsc{R{\'e}sum{\'e}.}
  Cet article est consacr{\'e} {\`a} l'{\'e}tude du spectre d'une certaine famille
  d'{\'e}quations de Schr{\"o}dinger quasi-p{\'e}riodiques sur l'axe r{\'e}el lorsque
  les courbes iso-{\'e}nerg{\'e}tiques adiabatiques sont non born{\'e}es dans la
  direction des moments. Dans des intevralles d'{\'e}nergies o{\`u} cette
  propri{\'e}t{\'e} est v{\'e}rifi{\'e}e, nous obtenons une formule asymptotique pour
  l'exposant de Lyapunov, et nous d{\'e}montrons que le spectre est
  purement singulier.
\end{abstract}
\setcounter{section}{-1}
\maketitle
\section{Introduction}
\label{sec:intro}
\begin{figure}
  \begin{center}
    \includegraphics[bbllx=71,bblly=591,bburx=569,bbury=721,height=4cm]{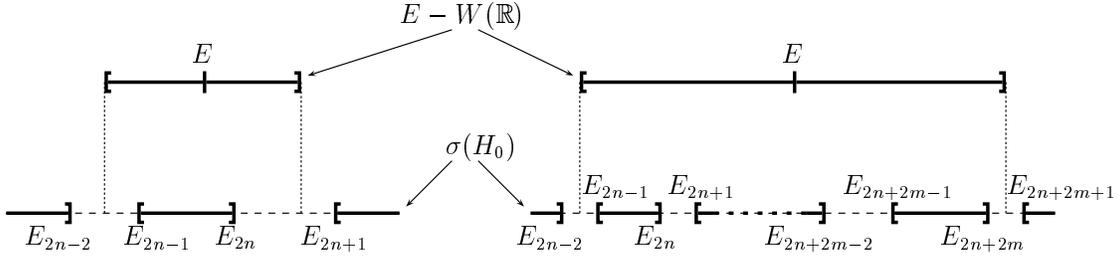}
  \end{center}
  \caption{The isolated band: two possible cases}\label{fig:ibm0}
\end{figure}

\noindent In this paper, we continue our analysis of the spectrum of
the ergodic family of Schr{\"o}dinger equations
\begin{equation}
  \label{family}
  H_{z,\varepsilon}\psi=-\frac{d^2}{dx^2}\psi(x)+
  (V(x-z)+W(\varepsilon x))\psi(x)= E\psi(x), \quad x\in\R,
\end{equation}
where $V(x)$ and $W(\xi)$ are periodic and real valued, $z\in\R$
indexes the equations, and $\varepsilon>0$ is chosen so that the
potential $V(\cdot-z)+W(\varepsilon\cdot)$ be quasi-periodic. We study
the spectral properties of the operator $H_{z,\varepsilon}$ acting in
$L^2(\R)$ in the limit as $\varepsilon\to0$. In the
paper~\cite{Fe-Kl:02}, we studied this operator near the bottom of the
spectrum when $W$ is the cosine. In the paper~\cite{Fe-Kl:01b}, for a
general analytic, periodic potential $W$, we studied the spectrum
located in the ``middle'' of a spectral band of the ``unperturbed''
periodic operator
\begin{equation}
  \label{Ho}
  H_0\psi(x)=-\psi''(x)+V(x)\psi(x).
\end{equation}
In the present paper, we again consider a rather general analytic
potential $W$; we only assume that it has exactly one maximum and one
minimum in a period, and that these are non-degenerate. As about $V$,
it can be rather singular; for the sake of simplicity, we assume that
it belongs to $L^2_{loc}$. We study the spectrum in an energy interval
$J$ such that, for all $E\in J$, the interval $E-W(\R)$ contains one
or more isolated spectral bands of the periodic operator~\eqref{Ho}
whereas the ends of the interval $E-W(\R)$ are in the gaps, see
Fig.~\ref{fig:ibm0}.  So, we are interested in the spectrum close to
and inside relatively small bands of the unperturbed periodic operator
$H_0$.
\smallpagebreak As in~\cite{Fe-Kl:02,Fe-Kl:01b}, our main tool is the
monodromy matrix. Most of the present paper is devoted to the
asymptotic study of the monodromy matrix for the family of
equations~\eqref{family}.  In the adiabatic limit $\varepsilon\to 0$,
the monodromy matrix is asymptotic to a trigonometric polynomial; if
the interval $E-W(\R)$ contains only one isolated spectral band, this
is a trigonometric polynomial of a first order. In result, the
analysis of~\eqref{family} reduces to the analysis of a ``simple''
model difference equation.
\smallpagebreak Using the monodromy matrix asymptotics, we obtain
asymptotic formulae for the Lyapunov exponent for the equation
family~\eqref{family}. They show that, in $J$, the energy region we
study, the Lyapunov exponent is positive. This implies that the
spectrum of~\eqref{family} in $J$ is singular.
\smallpagebreak The spectral results admit a natural semi-classical
interpretation.  Let $\mathcal{E}(\kappa)$ be the dispersion relation
associated to $H_0$. Consider the {\it real} and the {\it complex
  iso-energy curves} $\Gamma_\R$ and $\Gamma$ defined by
\begin{gather}
  \label{isoenr}
  \Gamma_\R:\quad\mathcal{E}(\kappa)+W(\zeta)=E,\quad \kappa,\zeta\in
  \R,\\
  \label{isoen}
  \Gamma:\quad\mathcal{E}(\kappa)+W(\zeta)=E,\quad \kappa,\zeta\in \C.
\end{gather}
These curves are $2\pi$-periodic as in $\zeta$ so in $\kappa$.  Under
our assumptions, the real branches of $\Gamma$ (the connected
components of $\Gamma_\R$) are isolated continuous curves periodic in
$\kappa$. In the case when the interval $E-W(\R)$ contains only one
spectral band, the iso-energy curve is shown in Fig.~\ref{fig:ibm4}.
The real branches are represented by full lines. They are connected by
complex loops (closed curves) lying on $\Gamma$; the loops are
represented by dashed lines.
%
%
%
%
\begin{floatingfigure}{6cm}
  \begin{center}
    \includegraphics[bbllx=71,bblly=566,bburx=247,bbury=721,width=6cm]{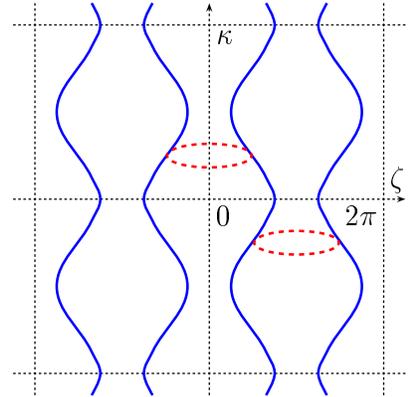}
  \end{center}
  \caption{The phase space picture}\label{fig:ibm4}
\end{floatingfigure}
%
\par The adiabatic limit can be regarded as a semi-classical limit, and
the expression $\mathcal{E}(\kappa)+W(\zeta)$ can be interpreted as a
``classical'' Hamiltonian corresponding to the
operator~\eqref{family}. Then, from the quantum physicist point of
view (see~\cite{Wi:84,MR88a:81040}), a semi-classical particle should
``live'' near the real branches of the iso-energy curve. In our case,
these curves are ``extended'' in momentum variable and ``localized''
in position variable. Therefore, they have to correspond to localized
states. The decay of these states in the position variable is
characterized by the complex tunneling between the real branches along
the complex loops. So, the Lyapunov exponent is naturally related to
the tunneling coefficients. Our results justify this heuristics.
\smallpagebreak This naturally leads to the following conjecture: in a
given energy interval, if the iso-energy curve has a real branch that
is an unbounded vertical curve, then, in the adiabatic limit, in this
interval, the Lyapunov exponent is positive and the spectrum is
singular.
\vskip.1cm\noindent Note that, in~\cite{Fe-Kl:01b}, we have proved a
dual result for the absolutely continuous spectrum: we have proved
that, if, in some energy region, the branches of the real iso-energy
curve are unbounded horizontal curves, then, this energy region,
except for a set of exponentially small measure, is in the absolutely
continuous spectrum.


%
\section{The results}
\label{sec:results}
\noindent We now state our assumptions and results.
\subsection{Assumptions on the potential}
\label{sec:assumpt-potent}
About the functions $V$ and $W$, we assume that
\begin{description}
\item[(H)]
  \begin{itemize}
  \item $V$ and $W$ are periodic,
    \begin{equation}\label{periodicity}
      V\,(x+1)=V\,(x),\quad W\,(x+2\pi)=W\,(x),\quad x\in\R;
    \end{equation}
  \item $V$ is real valued and locally square integrable;
  \item $W$ is real analytic in a neighborhood of $\R$, say, in the
    strip $\{|\im z|<Y\}$;
  \item $W$ has exactly one maximum and one minimum in $[0,2\pi)$;
    they are non degenerate.
  \end{itemize}
\end{description}
\smallpagebreak To fix notations, assume that, on the interval
$[0,2\pi)$, $W$ is maximum at $0$ and minimum at $\zeta^*$.
\smallpagebreak In~\eqref{family}, $\varepsilon$ is a positive
parameter. For each fixed $\varepsilon$, we consider~\eqref{family} as
a family of equations indexed by the parameter $z\in\R$.
\smallpagebreak Note that, if $2\pi/\varepsilon\not\in\Q$, the
function $V(x-z)+W(\varepsilon x)$ is quasi-periodic in $x$, the ratio
of the frequencies of $V$ and $W$ being equal to $2\pi/\varepsilon$;
hence,~\eqref{family} is an ergodic family of equations
(see~\cite{Pa-Fi:92}).
\subsection{The assumption on the energy region}
\label{SS:PSO1}
To describe the energy regions where we study the spectral properties
of the family of equations~\eqref{family}, we consider the periodic
Schr{\"o}dinger operator $H_0$ acting in $L^2(\R)$ defined by~\eqref{Ho}.
\subsubsection{Periodic operator}
\label{sec:periodic-operator}
The spectrum of~\eqref{Ho} is absolutely continuous and consists of
intervals of the real axis $[E_1,\,E_2]$, $[E_3,\,E_4]$, $\dots$,
$[E_{2n+1},\,E_{2n+2}]$, $\dots$, such that
\begin{gather*}
  E_1<E_2\le E_3<E_4\dots E_{2n}\le E_{2n+1}<E_{2n+2}\le \dots\,,\\
  E_n\to+\infty,\quad n\to+\infty.
\end{gather*}
The points $E_{j}$, \ $j=1,2,3\dots$, are the eigenvalues of the
differential operator~\eqref{Ho} acting on $L^2([0,2])$ with periodic
boundary conditions. The intervals defined above are called the {\it
  spectral bands}, and the intervals $(E_2,\,E_3)$, $(E_4,\,E_5)$,
$\dots$, $(E_{2n},\,E_{2n+1})$, $\dots$, are called the {\it spectral
  gaps}. If $E_{2n}<E_{2n+1}$, we say that the $n$th gap is {\it
  open}, and, if $[E_{2n-1},E_{2n}]$ is separated from the rest of the
spectrum by open gaps, we say that the $n$-th band is {\it isolated}.
\subsubsection{The ``geometric'' assumption}
\label{sec:main-assumption-w}
Let us now describe the energy region where we study the family of
equations~\eqref{family}.
\smallpagebreak The {\it spectral window} centered at $E$ is the
interval $\Wc(E)=E-W(\R)$. If $W_+=\max_{x\in\R} W(x)$ and
$W_-=\min_{x\in\R}W(x)$, then, $\Wc(E)=[E-W_+,E-W_-]$.
\smallpagebreak We assume that there exists $J\subset \R$, a compact
interval such that, for all $E\in J$, the window $\Wc(E)$ contains
exactly $m+1$ isolated bands of the periodic operator. That is, we fix
two integers $n>0$ and $m$ and assume that
\begin{description}
  \label{IBMcondition}
\item[(A1)] the bands $[E_{2(n+j)-1},E_{2(n+j))}]$, $j=0,1,\dots m$,
  are isolated;
\item[(A2)] for all $E\in J$, these bands are contained in the
  interior of $\Wc(E)$;
\item[(A3)] for all $E\in J$, the rest of the spectrum of the periodic
  operator is outside $\Wc(E)$.
\end{description}
Note that energies $E$ satisfying (A1) -- (A3) exist only if $W_+-W_-$,
the ``amplitude'' of the adiabatic perturbation, is large enough;
e.g., if $m=0$, such energies exist if and only if $W_+-W_-$ is larger
than the size of the $n$-th spectral band, but smaller than the
distance between the $(n-1)$-st and $(n+1)$-st bands.\\
From now on, unless stated otherwise, we assume that our assumptions
on $V$ and $W$, and assumptions (A1) -- (A3) are satisfied.
\subsection{Iso-energy curve}
\label{res:isoen}
Our results are formulated in terms of the iso-energy curve $\Gamma$
defined by~\eqref{isoen}. The iso-energy curve is $2\pi$ periodic both
in the $\zeta-$ and $\kappa-$ directions (see Lemma~\ref{Gamma:sym}).
\subsubsection{The real branches}
\label{res:real-branches}
To describe the real branches of $\Gamma$, i.e. the connected
components of the real iso-energy curve $\Gamma_\R$, we define the
following collection of subintervals of $[0,2\pi]$. Consider the
mapping
\begin{equation*}
  \mathcal{E}:\zeta\to E-W(\zeta).
\end{equation*}
It is monotonous on each of the intervals $I_-=[0,\zeta^*]$ and
$I_+=[\zeta^*,2\pi]$ and maps each of them onto the spectral window
$\Wc(E)$. For $n\leq j\leq n+m$, let $\mathfrak{z}_j^+\subset I_+$
(resp. $\mathfrak{z}_j^-\subset I_-$) be the the pre-image of the
$j$-th spectral band in $\Wc(E)$. Let $\mathcal Z$ be the collection
of these intervals. A ``period'' of the real iso-energy curve is
described by
\begin{Le}
  \label{realbranches}
  Let $E\in J$.  The set $\Gamma_\R\cap\{0\le\kappa\le2\pi\}$ consists
  of $2(m+1)$ curves $\{\gamma(\mathfrak{z})$,
  $\mathfrak{z}\in\mathcal{Z}\}$.\\
  Fix $\mathfrak{z}\in{\mathcal Z}$. The curve $\gamma(\mathfrak{z})$
  is the graph $\{(\kappa,\zeta):\ \zeta=Z_{\mathfrak{z}}(\kappa), \ 
  \kappa\in\R\}$ of a function $Z_{\mathfrak{z}}$ which satisfies
  \begin{enumerate}
  \item it is continuous,
  \item it is $2\pi$-periodic and even in $\kappa$,
  \item it is monotonous on the interval $[0,\pi]$,
  \item it maps $[0,\pi]$ onto $\mathfrak z$.
  \end{enumerate}
  The curves $\gamma(\mathfrak{z})$ continuously depend on $E\in J$.
\end{Le}
\noindent Lemma~\ref{realbranches} is proved in
section~\ref{GammaR}. For $m=0$, the real iso-energy curve is shown in
Fig.~\ref{fig:ibm4}.
\subsubsection{Complex loops}
\label{res:cl-cur}
\begin{figure}[ht]
  \begin{center}
    \includegraphics[bbllx=71,bblly=634,bburx=460,bbury=721,width=13.5cm]{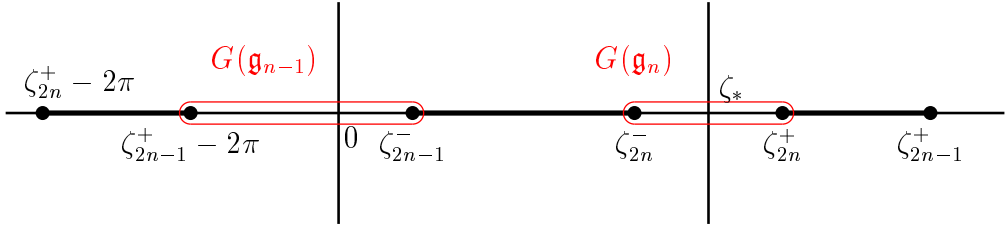}
  \end{center}
  \caption{The curves $G(\mathfrak g)$ for $m=0$}\label{IBMfig:actions}
\end{figure}

Now, we discuss loops, i.e. closed curves, situated on the iso-energy
curve $\Gamma$ and connecting its real branches.
\smallpagebreak For $j=n-1,n,\dots,n+m$, let $\mathfrak{g}_j^+$
(resp. $\mathfrak{g}_j^-$) be the subinterval of $I_+$ (resp. $I_-$)
that is the pre-image of the part of $j$-th spectral gap situated
inside $\Wc(E)$. Let
\begin{equation*}
  \mathfrak{g}_{n-1}=(\mathfrak{g}_{n-1}^+-2\pi)\cup
  \mathfrak{g}_{n-1}^-\quad \text{and} \quad
  \mathfrak{g}_{n+m}=\mathfrak{g}_{n+m}^+\cup\mathfrak{g}_{n+m}^-.
\end{equation*}
Then, $\mathfrak{g}_{n-1}$ is an open interval containing zero, and
$\mathfrak{g}_{n+m}$ is an open interval containing $\zeta^*$. Let
$\mathcal G$ be the set consisting of $g_{n-1}$, $g_{n+m}$ and the
intervals $g_{j}^\pm$ with $j=n,n+1,\dots,n+m-1$.
\smallpagebreak For $\mathfrak{g}\in{\mathcal G}$, let
$V(\mathfrak{g})\subset \C$ be a sufficiently small complex
neighborhood of the interval $\mathfrak{g}$. Let $G(\mathfrak{g})$ be
a smooth closed curve that goes once around the interval
$\mathfrak{g}$ in $V(\mathfrak{g})\setminus \mathfrak g$. In
Figure~\ref{IBMfig:actions}, we depicted the curves $G(\mathfrak{g})$
when $m=0$.
\smallpagebreak In section~\ref{s:cl-cur}, we show that each of the
curves $G(\mathfrak{g})$ is the projection of $\hat G(\mathfrak{g})$,
a closed curve on $\Gamma$. This curve connects the real branches
projecting onto the intervals adjacent to $\mathfrak g$.
\subsubsection{Tunneling coefficients}
\label{sec:tunn-coeff}
To $\Gamma$, we associate the tunneling coefficients
\begin{equation}
  \label{t(g)}
  t(\mathfrak{g})=e^{\dsize-\frac1{2\varepsilon} S(\mathfrak{g})},\quad
  \mathfrak{g}\in{\mathcal G},
\end{equation}
where $S(\mathfrak{g})$ are the {\it tunneling actions} given by
\begin{equation}
  \label{S(g)}
  S(\mathfrak{g})=i\oint_{\hat G(\mathfrak{g})}\kappa d\zeta,\quad
  \mathfrak{g}\in{\mathcal G}.
\end{equation}
In section~\ref{t-coeff}, we show that, for $E\in J$, each of these
actions is real and non-zero. By definition, we choose the direction
of the integration so that all the tunneling actions be positive.
%
%
%
\subsection{Spectral results}
\label{sec:spectral-results}
One of the main objects of the spectral theory of quasi-periodic
equations is the Lyapunov exponent, see, for example,~\cite{Pa-Fi:92}.
Our main spectral result is
\begin{Th}
  \label{th:ibm:sp} Let $J$ be an interval satisfying the assumptions
  (A1)-(A3) for some $n$ and $m$. Let $W$ and $V$ satisfy the
  hypothesis (H), and let $\varepsilon$ be irrational. Then, on the
  interval $J$, for sufficiently small, irrational $\varepsilon/2\pi$,
  the Lyapunov exponent $\Theta(E)$ for the family of
  equations~\eqref{family} is positive and has the asymptotics
  \begin{equation}
    \label{LexpA}
    \Theta(E)=\frac{\varepsilon}{2\pi}\sum_{\mathfrak{g}\in{\mathcal G}}
    \ln\frac1{t(\mathfrak{g})}+o(1)=
    \frac{1}{4\pi}\sum_{\mathfrak{g}\in{\mathcal G}} S(\mathfrak{g})+o(1).
  \end{equation}
\end{Th}
\noindent Note that, this theorem implies that, if $\varepsilon$ is
sufficiently small, then, the Lyapunov exponent is positive for all
$E\in J$.
\smallpagebreak Recall that, if $2\pi/\varepsilon$ is irrational,
then $H_{z,\varepsilon}$ is quasi-periodic. In this case, its
spectrum does not depend on $z$ (see~\cite{Av-Si:83}); denote it
by $\sigma(H_{z,\varepsilon})$. In~\cite{Fe-Kl:02}, we have proved
\begin{Th}[\cite{Fe-Kl:02}]
  \label{thr:3}
  Let $\Sigma=\sigma(H_0)+W(\R)=\sigma(H_0)+[W_-,W_+]$. Then, one has
  \begin{itemize}
  \item $\forall\varepsilon\geq0$, $\sigma(H_{z,\varepsilon})\subset\Sigma$.
  \item for any $K\subset\Sigma$ compact, there exists $C>0$ such
  that for all $\varepsilon$ sufficiently small and $\forall E\in K$, one has
    \begin{equation*}
     \sigma(H_{z,\varepsilon})\cap(E-C\varepsilon^{1/2},E+C\varepsilon^{1/2})
    \not=\emptyset.
    \end{equation*}
  \end{itemize}
\end{Th}
\noindent By  the Ishii-Pastur-Kotani
Theorem~\cite{Cy-Fr-Ki-Si:87,Pa-Fi:92} and Theorem 1.5
in~\cite{MR2000f:47060}, Theorems~\ref{th:ibm:sp} and~\ref{thr:3}
imply
\begin{Cor}
  \label{cor:1}
  In the case of Theorem~\ref{th:ibm:sp}, for $\varepsilon$
  sufficiently small, for all $z\in\R$, one has
  \begin{equation*}
    \sigma(H_{z,\varepsilon})\cap J\not=\emptyset\quad\text{ and
    }\quad\sigma_{ac}(H_{z,\varepsilon})\cap J=\emptyset,
  \end{equation*}
  where $\sigma_{ac}(H_{z,\varepsilon})$ is the absolutely continuous
  spectrum of the family of equations~\eqref{family}.
\end{Cor}
\subsection{The monodromy matrix and Lyapunov exponents}
\label{sec:monodromy-matrix-1}
The main object of our study is the monodromy matrix for the family of
equations~\eqref{family}; we define it briefly (we refer
to~\cite{Fe-Kl:01a,Fe-Kl:02} for more details). The central result of
the paper is its asymptotics in the adiabatic limit.
\subsubsection{Definition of the monodromy matrix}
\label{sec:defin-monodr-matr}
Consider a consistent basis $(\psi_{1,2})$ i.e. a basis of solutions
of~\eqref{family} whose Wronskian is independent of $z$ and that are
$1$-periodic in $z$ i.e. that satisfy
\begin{equation}\label{consistency}
  \psi_{1,2}(x,\,z+1)=\psi_{1,2}(x,\,z),\quad \forall x,z.
\end{equation}
The functions $\psi_{1,2}(x+2\pi/\varepsilon,z+2\pi/\varepsilon)$
being solutions of equation~\eqref{family}, one can write
\begin{equation}\label{monodromy}
  \Psi\,(x+2\pi/\varepsilon,z+2\pi/\varepsilon)= M\,(z)\,
\Psi\,(x,z),
\end{equation}
where
\begin{itemize}
\item $\D\Psi(x,z)=\begin{pmatrix}\psi_{1}(x,\,z)\\
    \psi_{2}(x,z)\end{pmatrix}$,
\item $M\,(z)$ is a $2\times 2$ matrix with coefficients independent
  of $x$.
\end{itemize}
The matrix $M$ is called {\it the monodromy matrix} associated to
the consistent basis $(\psi_{1,2})$. Note that
\begin{equation}\label{Mproperties}
  \det M\,(z)\equiv 1,\quad
  M\,(z+1)=M\,(z),\quad \forall z.
\end{equation}
\subsubsection{Monodromy equation and Lyapunov exponents}
\label{sec::Mon-eq}
Set $h=\frac{2\pi}{\varepsilon}\,{\rm mod}\, 1$. Let $M$ be the
monodromy matrix associated to a consistent basis $(\psi_{1,2})$.
Consider the {\it monodromy equation}
\begin{equation}
  \label{Mequation}
  F_{n+1}=M(z+nh)F_n\quad\quad \forall n\in\Z.
\end{equation}
\smallpagebreak There are several deep relations between the monodromy
equation and the family of equations~\eqref{family}
(see~\cite{Fe-Kl:01b,Fe-Kl:02}).  We describe only one of them.  Let
$2\pi/\varepsilon$ be irrational, and let $\Theta(E)$ (resp.
$\theta(E)$) be the Lyapunov exponent for~\eqref{family} (resp.
for~\eqref{Mequation}). One proves
\begin{Th}[\cite{Fe-Kl:02}]
  \label{Lyapunov}
  The Lyapunov exponents $\Theta(E)$ and $\theta(E)$ satisfy the
  relation
  \begin{equation}\label{eq:Lyapunov}
    \Theta(E)=\frac\varepsilon{2\pi}\theta(E).
  \end{equation}
\end{Th}
\smallpagebreak The passage to the monodromy equation is close to the
monodromization idea developed in~\cite{Bu-Fe:96} for difference
equations with periodic coefficients.
\subsubsection{The asymptotics of the monodromy matrix}
\label{sec:monodromy-matrix}
As $W$ and $V$ are real on the real line, we construct a monodromy
matrix of the form
\begin{equation}
  \label{Mform}
  \begin{pmatrix} a(z,E) & b(z,E)\\ \overline{b(\bar z,\bar E)} &
    \overline{a(\bar z,\bar E)}\end{pmatrix}.
\end{equation}
In the adiabatic case, the asymptotics of $a$ and $b$ have very
simple, model form. We first assume that $n$ in (A1) -- (A3) is odd.
Then, one has
\begin{Th}
  \label{Th:mat-mon:1}
  Let $E\in J$. There exists $Y>0$ and $V_0$, a neighborhood of $E_0$,
  such that, for sufficiently small $\varepsilon$, the family of
  equations~\eqref{family} has a consistent basis of solutions for
  which the corresponding monodromy matrix $M$ is analytic in
  $(z,E)\in\{|\im z|<Y/\varepsilon\}\times V_0$ and has the
  form~\eqref{Mform}. The coefficients $a$ and $b$ admit the
  asymptotic representations
  \begin{equation}\label{a,b:up}
    a=   a_{-m}\,e^{-2\pi i m z}(1+o(1)),\quad
    b=   b_{-m}\, e^{-2\pi i m z}(1+o(1)),\quad 0<\im z<Y/\varepsilon,
  \end{equation}
  and
  \begin{equation}\label{a,b:down}
    a=   a_{m+1}\, e^{ 2\pi i (m+1) z}(1+o(1)),\quad
    b=   b_{m+1}\, e^{ 2\pi i (m+1) z}(1+o(1)),\quad
    -Y/\varepsilon<\im z<0.
  \end{equation}
  The coefficients $a_{-m}$, $b_{-m}$, $a_{m+1}$ and $b_{m+1}$ are
  independent of $z$. Moreover, there exists a constant $C>1$
  (independent of $\varepsilon$ and $E$) such that
  \begin{equation}\label{Fm}
  \begin{array}{c}
    \dsize\frac1C\le T\cdot |a_{j}|\le C,\quad\quad
    \frac1C\le T\cdot|b_{j}|\le C,\quad\text{where}\quad
      T=T(E)=\prod_{\mathfrak{g}\in{\mathcal G}} t(\mathfrak{g}),\\{}\\
     E\in V_0\cap \R,\quad  j=-m,\ m+1.
  \end{array}
  \end{equation}
  Pick $Y_1$ and $Y_2$ so that $0<Y_1<Y_2<Y$.  There is
  $V=V(Y_1,Y_2)$, a neighborhood of $E_0$ such that the asymptotics of
  $a$ and $b$ are uniform in $(z,E)\in \{Y_1<|\im\zeta|<Y_2\}\times
  V$.
\end{Th}
\smallpagebreak In sections~\ref{b:odd} and~\ref{a:odd}, we give
asymptotic formulae for $a_{m+1}$, $a_{-m}$ and $b_{m+1}$, $b_{-m}$.
\smallpagebreak In the case $n$ even, one has a similar result. The
only novelty is that, in this case, the formulae~\eqref{a,b:up}
and~\eqref{a,b:down} describe the asymptotics of the coefficients of
the matrix related to $M$, the monodromy matrix, by the following
transformation
\begin{equation}
  \label{SMS}
  S^{-1}(z+h)M(E,z) S(z),\quad  S(z)=\begin{pmatrix} e^{i\pi z}& 0\\ 0 &
    e^{-i\pi z}\end{pmatrix}.
\end{equation}
The asymptotics~\eqref{a,b:up} and~\eqref{a,b:down} are obtained by
means of the new asymptotic method developed
in~\cite{Fe-Kl:01a,Fe-Kl:03a}.
\subsubsection{Fourier coefficients}
\label{sec:fourier-coefficients}
The coefficients $a_{m+1}$, $a_{-m}$ and $b_{m+1}$, $b_{-m}$ are the
leading terms of the asymptotics of the $(m+1)$-th and $(-m)$-th
Fourier coefficients of the monodromy matrix coefficients.
Theorem~\ref{Th:mat-mon:1} implies that, in the strip
$\{|\im\zeta|<Y\}$, the leading terms of the asymptotics of the
monodromy matrix are equal to the contribution of a few of its Fourier
series terms.
\subsubsection{The case $m=0$}
\label{sec:case-m=0}
When $m=0$ (and $n$ odd), Theorem~\ref{Th:mat-mon:1} imply that, in
the whole strip $\{|\im z|<Y\}$, the monodromy matrix coefficients $a$
and $b$ admit the asymptotics:
\begin{equation}\label{a,b:m=1}
  a= a_{0}\, (1+o(1))+a_1\,e^{2\pi i  z}(1+o(1)),\quad
  b=   b_{0}\,(1+o(1))+b_1\,e^{2\pi i z}(1+o(1)).
\end{equation}
So, up to the error terms, the monodromy matrix becomes a first
order trigonometric polynomial:
\begin{equation}
  \label{M_0}
  M\sim M_0=\begin{pmatrix} a_0+a_1 u & b_0+b_1 u \\
  \overline{b_0}+\overline{b_1}/u & \overline{a_0}+\overline{a_1}/u
  \end{pmatrix}, \quad u=e^{2\pi i z},
\end{equation}
with constant coefficients $a_0$, $a_1$, $b_0$, $b_1$ of order
$O(1/T(E))$ (for real $E$).
\smallpagebreak We see that, for $m=0$ the monodromy equation becomes
a ``simple'' model equation.
\subsubsection{Relation to the spectral results}
\label{sec:relat-spectr-results}
In this paper, we use the asymptotics of the monodromy matrix only to
prove Theorem~\ref{th:ibm:sp}. However, we believe that these
asymptotics can be used to get quite a detailed information on the
spectrum of~\eqref{family} in the adiabatic limit. Therefore, we plan
to study the model equation with the matrix $M_0$ in a subsequent
paper. In particular, it seems reasonable to believe that, under a
Diophantine condition on $2\pi/\varepsilon$, the spectrum
of~\eqref{family} is pure point and the eigenvalues can be described
by quantization conditions of Bohr-Sommerfeld type.
\subsubsection{Organization of the paper}
\label{sec:organization-paper}
Section~\ref{sec:pos-lyap} is devoted to the proof of
Theorem~\ref{th:ibm:sp} using Theorem~\ref{Th:mat-mon:1}. In
section~\ref{S3}, we recall some well known facts from the theory of
periodic Schr{\"o}dinger operators on the real line. In section~\ref{S4},
we recall the main construction of the asymptotic method we use to
compute the monodromy matrix. In sections~\ref{sec:f} and~\ref{basis},
using this method, we construct a consistent basis of solutions having
a simple ``standard'' asymptotic behavior in the complex plane of
$\zeta=\varepsilon z$. In section~\ref{sec:m-m:general}, we discuss
the properties of the monodromy matrix for this basis. This is the
monodromy matrix the asymptotics of which are described in
Theorem~\ref{Th:mat-mon:1}.  Sections~\ref{wronskians:gen-as}
and~\ref{M-asymptotics} are devoted to the computation of the
asymptotics of the monodromy matrix.  In section~\ref{sec:curves}, we
study the geometry of the iso-energy curve $\Gamma$ and prove
estimates~\eqref{Fm}.


%
\section{The asymptotics for the Lyapunov exponent}
\label{sec:pos-lyap}
\noindent In this section, we prove the asymptotics~\eqref{LexpA}. We
deduce these asymptotics from the asymptotics of the monodromy matrix
coefficients described by Theorem~\ref{Th:mat-mon:1}. First, we use a
statement of~\cite{Fe-Kl:02} and obtain a lower bound for the Lyapunov
exponent. This statement is based on the ideas of~\cite{MR93b:81058}
generalizing Herman's argument~\cite{MR85g:58057}. Then, using the
asymptotics of the coefficients of the monodromy matrix in the complex
plane, we get estimates on the real line. This yields an upper bound
for the Lyapunov exponent. Comparing the upper and the lower bounds,
we obtain~\eqref{LexpA}.
\smallpagebreak Recall that, for equation~\eqref{Mequation}, the
Lyapunov exponent is defined by
\begin{equation}
  \label{eq:44}
  \theta(M)=\lim_{N\to+\infty}\frac{1}{N} \log\Vert
  P_N(z)\Vert,
\end{equation}
where $P_N$ is the matrix cocycle
\begin{equation*}
  P_N(z)= M(z+Nh)\cdot M(z+(N-1)h)\cdots
  M(z+h)\cdot M(z).
\end{equation*}
It is well known (see~\cite{Ca-La:90,Pa-Fi:92,MR93b:81058} and
references therein) that, if $h$ is irrational, and $M(z)$
sufficiently regular in $z$, then the limit~\eqref{eq:44} exists for
almost all $z$ and is independent of $z$.
\subsection{The lower bound}
\label{sec:lower-bound}
\subsubsection{Preliminaries}
\label{sec:preliminaries}
Let $(M(z,\varepsilon))_{0<\varepsilon<1}$ be a family of
$SL(2,\C)$-valued $1$-periodic functions of $z\in\C$. Let $h$ be an
irrational number. One has
\begin{Pro}[\cite{Fe-Kl:02}]
  \label{th:lyap>0} 
  Pick $\varepsilon_0>0$. Assume that there exist $y_{0}$ and $y_{1}$
  satisfying the inequalities $0<y_0<y_1<\infty$ and such that, for
  any $\varepsilon\in(0,\varepsilon_0)$ one has
  \begin{itemize}
  \item the function $z\to M(z,\varepsilon)$ is analytic in the strip
    $S=\{z\in\C;\ 0\le \im z\le y_1/\varepsilon\}$;
  \item in the strip $S_{1}=\{z\in\C;\ y_0/\varepsilon\le\im z \le
    y_1/\varepsilon \}\subset S$, $M(z,\varepsilon)$ admits the
    representation
    \begin{equation*}
      M(z,\varepsilon)=\lambda(\varepsilon)e^{i2\pi n_{0} z}\cdot
      \left(M_0(\varepsilon)+M_1(z,\varepsilon)\right),
    \end{equation*}
    for some constant $\lambda(\varepsilon)$, an integer $n_0$ and a
    matrix $M_{0}(\varepsilon)$, all of them independent of $z$;
  \item $M_0(\varepsilon)=\begin{pmatrix}1&\beta(\varepsilon)\\0
      &\alpha(\varepsilon)\end{pmatrix}$;
  \item there exist constants $\beta>0$ and $\alpha \in(0,1)$
    independent of $\varepsilon$ and such that
    $|\alpha(\varepsilon)|\leq\alpha$ and
    $|\beta(\varepsilon)|\leq\beta$;
  \item $\sup_{z\in S_{1}}\Vert M_1(z,\varepsilon)\Vert\le
    m(\varepsilon)$, \ $m(\varepsilon)\to 0$ as $\varepsilon\to 0$.
  \end{itemize}
  Then, there exit $C>0$ and $\varepsilon_1>0$ (both depending only on
  $y_0$, $y_1$, $\alpha$, $\beta$ and $m(\cdot )$) such that, if
  $0<\varepsilon<\varepsilon_1$, one has
  \begin{equation}
    \label{eq:l>0}
    \theta(M)>\log|\lambda(\varepsilon)|-Cm(\varepsilon).
  \end{equation}
\end{Pro}
\smallpagebreak In~\cite{Fe-Kl:02}, we have assumed that $n_0$ is a
positive integer, but the proof remains the same for $n_0\in\Z$.
\smallpagebreak We use this result and Theorem~\ref{Th:mat-mon:1} to
get the lower bound for the Lyapunov exponent. In the sequel, $n$ is
the index introduced in assumptions (A1) -- (A3). The cases $n$ odd
and $n$ even are treated separately.
\subsubsection{Obtaining the lower bound for $n$ odd}
\label{lb:odd}
In the sequel, we suppose that the assumptions of
Theorem~\ref{Th:mat-mon:1} are satisfied; we use the notations and
results from this theorem without referring to it anymore. We assume
that $E\in V_0\cap\R$.\\
Let $\sigma=\begin{pmatrix} 0 & 1 \\ 1 & 0 \end{pmatrix}$. Show that
the matrix $\sigma M(z)\sigma$ satisfies the assumptions of
Proposition~\ref{th:lyap>0}.\\
Fix $y_0$ and $y_1$ so that $0<y_0<y_1<Y$. The asymptotics of the
monodromy matrix coefficients are uniform in $z$ in the strip
$S=\{y_0/\varepsilon\le \im z\le y_1/\varepsilon\}$ and in $E\in V_0$
(reducing $V_0$ if necessary). For $E\in V_0\cap \R$ and $z\in S$,
formulae~\eqref{a,b:up} and~\eqref{a,b:down}, and estimates~\eqref{Fm}
imply that
\begin{equation*}
  \overline{a({\overline z})}= \overline{a_{m+1}} 
  e^{-2\pi i (m+1) z}(1+o(1)),
  \frac{\overline{b({\overline z})}}{\overline{a({\overline z})}}
  = c(E)\, (1+o(1)),
  \frac{a(z)}{\overline{a({\overline z})}}=o(1),\quad
  \frac{b}{\overline{a({\overline z})}}=o(1)
\end{equation*}
where $c(E)$ is independent of $z$ and bounded by a constant uniformly
in $\varepsilon$ and $E$. So, we have
\begin{equation*}
  \sigma\,M(z)\sigma=\overline{a_{m+1}}\,e^{-2\pi i(m+1)z}\,\left[
    \begin{pmatrix}    1 & c(E) \\ 0 & 0 \end{pmatrix}+ o(1)\right].
\end{equation*}
We see that the matrix $\sigma M(z)\sigma$ satisfies the assumptions
of Proposition~\ref{th:lyap>0}. \\ 
Clearly, the Lyapunov exponents of the matrix cocycles associated to
the pairs $(M,h)$ and $(\sigma M\sigma, h)$ coincide. So,
Proposition~\ref{th:lyap>0} implies that $\theta(M)$, the Lyapunov
exponents of the matrix cocycle associated to $(M,h)$, satisfies the
estimate $\theta(M)\geq \log |a_{-m}|+o(1)$.\\
The Lyapunov exponent $\Theta(E)$ for equation~\eqref{family} is
related to $\theta(M)$ by Theorem~\ref{Lyapunov}. Therefore,
$\Theta(E)\ge\frac\varepsilon{2\pi}\log |a_{-m}|+o(\varepsilon)$.
Hence,~\eqref{Fm} clearly implies
\begin{equation}
  \label{lb}
 \Theta(E)\ge \frac\varepsilon{2\pi}\log T^{-1}+O(\varepsilon).
\end{equation}
\subsubsection{The lower bound when $n$ is even}
\label{sec:lower-bound-even}
If $n$ is even, then, formulae~\eqref{a,b:up} and~\eqref{a,b:down}
give the asymptotics of the coefficients of the matrix~\eqref{SMS}.
Obviously, the Lyapunov exponents for the matrix cocycles generated by
$M(z)$ and by $S^{-1}(z+h)M(z) S(z)$ coincide. Arguing exactly as in
subsection~\ref{lb:odd}, we again obtain~\eqref{lb}.
\subsection{The upper bound}
\label{sec:upper-bound}
Let us first assume that $n$ is odd. Let $E\in V_0\cap\R$. Fix
$0<y_0<Y$. The asymptotics~\eqref{a,b:up} and \eqref{a,b:down} and
estimates~\eqref{Fm} imply the following estimates for the
coefficients of $M(z)$, the monodromy matrix:
\begin{equation}\label{ub:1}
  \begin{array}{c}
    |a|,|b|\leq C(y_0) T^{-1}e^{2\pi m y_0/\varepsilon},\quad\im
    z=y_0/\varepsilon,\\
    |a|,|b|\leq C(y_0) T^{-1} e^{2\pi (m+1) y_0/\varepsilon},\quad
    \im z=-y_0/\varepsilon.
  \end{array}
\end{equation}
Here, $C(y_0)$ is a positive constant independent of $\varepsilon$,
$\re z$, and $E$. The estimates are valid for sufficiently small
$\varepsilon$. Recall that $M$ is analytic and $1$-periodic in $z$.
Therefore,~\eqref{ub:1} and the Maximum Principle imply that
\begin{equation}
  \label{ub:2}
  |a|,|b|\le 2C(y_0) T^{-1}
  \exp(2\pi (m+1)y_0/\varepsilon),\quad z\in\R.
\end{equation}
This leads to the following upper bound for the Lyapunov exponent for
the matrix cocycle generated by $M$
\begin{equation*}
  \theta\le \log T^{-1}+ \Const+2\pi (m+1) y_0/\varepsilon  
\end{equation*}
where $\Const$ is independent of $E$ and $\varepsilon$. In view of
Theorem~\ref{Lyapunov}, we finally get
\begin{equation}
  \label{ub}
  \Theta(E)\leq\frac\varepsilon{2\pi}\log T^{-1}+
  \varepsilon \Const+ 2\pi (m+1) y_0.
\end{equation}
\smallpagebreak The upper bound~\eqref{ub} remains true when $n$ is
even as the Lyapunov exponents for the matrix cocycles generated by
$M(z)$ and by $S^{-1}(z+h)M(z)S(z)$ coincide.
\subsection{Completing the proof}
\label{sec:completing-proof}
Recall that, in~\eqref{ub}, $y_0$ is an arbitrarily fixed positive
number. So, comparing~\eqref{lb} and~\eqref{ub}, we get
$\D\Theta(E)=\frac\varepsilon{2\pi}\log T^{-1}+o(1)$. This and the
formula~\eqref{Fm} for $T$ imply~\eqref{LexpA} for all $E\in
V_0\cap\R$.\\
Recall that $V_0\cap\R$ is an open interval containing $E_0\in J$.
The above construction can be carried out for any $E_0\in J$.  As the
interval $J$ is compact, this completes the proof of
Theorem~\ref{th:ibm:sp}.


%
\section{Periodic Schr{\"o}dinger operators}
\label{S3}
\noindent We now discuss the periodic Schr{\"o}dinger operator~\eqref{Ho}
where $V$ is a $1$-periodic, real valued, $L^2_{loc}$-function. We
collect known results needed in the present paper
(see~\cite{Eas:73,Ma-Os:75,McK-Tr:75,Ti:58,Fe-Kl:03a}).
\subsection{Bloch solutions}
\label{SSS:BS}
Let $\psi$ be a solution of the equation
\begin{equation}\label{PSE}
  -\frac{d^2}{dx^2}\psi\,(x)+ V\,(x)\psi\,(x)=\mathcal{E}\psi\,(x),
   \quad x\in\R,
\end{equation}
satisfying the relation $\psi\,(x+1)=\lambda\,\psi\,(x)$ for all
$x\in\R$ with $\lambda\in\C$ independent of $x$. Such a solution is
called a {\it Bloch solution}, and the number $\lambda$ is called the
{\it Floquet multiplier}. Let us discuss the analytic properties of
$\mathcal{E}\mapsto\psi(\mathcal{E},x):=\psi(x)$.
\smallpagebreak As in section~\ref{SS:PSO1}, we denote the spectral
bands of the periodic Schr{\"o}dinger equation by $[E_1,\,E_2]$,
$[E_3,\,E_4]$, $\dots$, $[E_{2n+1},\,E_{2n+2}]$, $\dots$. Consider
$\mathcal{S}_\pm $, two copies of the complex plane $\mathcal{E}\in\C$
cut along the spectral bands. Paste them together to get a Riemann
surface with square root branch points.  We denote this Riemann
surface by $\mathcal{S}$.
\smallpagebreak One can construct a Bloch solution
$\psi(x,\mathcal{E})$ of equation~\eqref{PSE} meromorphic on $\mathcal
S$. It is normalized by the condition $\psi(1,\mathcal{E})\equiv 1$.
The poles of this solution are located in the spectral gaps. More
precisely, each spectral gap contains precisely one simple pole. It is
located either on $\mathcal{S}_+$ or on $\mathcal{S}_-$. The position
of the pole is independent of $x$.
\smallpagebreak For $\mathcal{E}\in\mathcal{S}$, we denote by
$\hat{\mathcal{E}}$ the point on $\mathcal{S}$ having the same
projection on $\C$ as $\mathcal{E}$. We let
\begin{equation*}
  \hat \psi(x,\mathcal{E})=\psi(x,\hat{\mathcal{E}}),\quad
  \mathcal{E}\in\mathcal{S}.
\end{equation*}
The function $\hat\psi(x,\mathcal{E})$ is another Bloch solution
of~\eqref{PSE}. Except at the edges of the spectrum (i.e. the branch
points of $\mathcal{S}$), the functions $\psi$ and $\hat\psi$ are
linearly independent solutions of~\eqref{PSE}. In the spectral gaps,
$\psi$ and $\hat\psi$ are real valued functions of $x$, and, on the
spectral bands, they differ only by complex conjugation.
\subsection{The Bloch quasi-momentum}
\label{SS3.2}
Consider the Bloch solution $\psi(x,\mathcal{E})$. The
corresponding Floquet multiplier $\lambda\,(\mathcal{E})$ is
analytic on $\mathcal{S}$. Represent it in the form
$\lambda(\mathcal{E})=\exp(ik(\mathcal{E}))$.
The function $k(\mathcal{E})$ is the {\it Bloch quasi-momentum}.
\smallpagebreak The Bloch quasi-momentum is an analytic
multi-valued function of $\mathcal{E}$. It has the same branch
points as $\psi(x,\mathcal{E})$.
\smallpagebreak Let $D$ be a simply connected domain containing no
branch point of the Bloch quasi-momentum. In $D$, one can fix an
analytic single-valued branch of $k$, say $k_0$. All the other
single-valued branches of $k$ that are analytic in $D$ are related to
$k_0$ by the formulae
\begin{equation}\label{eq:55}
   k_{\pm ,l}(\mathcal{E})=\pm k_0(\mathcal{E})+2\pi l,\quad l\in\Z.
\end{equation}
\smallpagebreak Consider $\C_+$ the upper half plane of the
complex plane. On $\C_+$, one can fix a single valued analytic
branch of the quasi-momentum continuous up to the real line. It
can be fixed uniquely  by the condition $ -ik(\mathcal{E}+i0)>0$
as $\mathcal{E}<E_1$. We call this branch the main branch of the
Bloch quasi-momentum and denote it by $k_p$.
\smallpagebreak The function $k_p$ conformally maps $\C_+$ onto the
first quadrant of the complex plane cut at compact vertical slits
starting at the points $\pi l$, $l\in \N$. It is monotonically
increasing along the spectral zones so that $[E_{2n-1}, E_{2n}]$, the
$n$-th spectral band, is mapped on the interval $[\pi(n-1), \pi n]$.
Along any open gap, $\re k_p(\mathcal{E}+i0)$ is constant, and $\im
k_p(\mathcal{E}+i0)$ is positive and has only one non-degenerate
maximum.
\smallpagebreak All the branch point of $k_{p}$ are of square root
type. Let $E_l$ be a branch point. In a sufficiently small
neighborhood of $E_l$, the function $k_p$ is analytic in
$\sqrt{\mathcal{E}-E_l}$, and
\begin{equation}
  \label{sqrt}
  k_{p}(\mathcal{E})-k_{p}(E_l)=c_l\sqrt{\mathcal{E}-E_l}
  +O(\mathcal{E}-E_l),\quad c_l\not=0.
\end{equation}
Finally, we note that the main branch can be analytically continued on
the complex plane cut only along the spectral gaps of the periodic
operator.
\subsection{A meromorphic function}
\label{sec:Omega}
Here, we discuss a function playing an important role in the adiabatic
constructions.
\smallpagebreak In~\cite{Fe-Kl:03a}, we have seen that, on $\mathcal
S$, there is a meromorphic function $\omega$ having the following
properties:
\begin{itemize}
\item the differential $\Omega=\omega\,d\mathcal{E}$ is meromorphic;
  its poles are the points of $P\cup Q$, where $P$ is the set of poles
  of $\psi(x,\mathcal{E})$, and $Q$ is the set of points where
  $k'(\mathcal{E})=0$.
\item all the poles of $\Omega$ are simple;
\item
  $\res_p \Omega=1$, \ $\forall p\in P\setminus Q$, \ \ 
  $\res_q\Omega=-1/2$, \ $\forall q\in Q\setminus P$, \ \ 
  $\res_r\Omega=1/2$, \ $\forall r\in P\cap Q$.
\item if $\mathcal{E}\in\mathcal S$ projects into a gap, then
  $\omega(\mathcal{E})\in \R$;
\item if $\mathcal{E}\in\mathcal S$ projects inside a band, then
  $\overline{\omega(\mathcal{E})}=\omega(\hat{\mathcal{E}})$.
\end{itemize}
%

%
\section{The complex WKB method for adiabatic problems}
\label{S4}
\noindent In this section, following~\cite{Fe-Kl:01a,Fe-Kl:03a}, we
briefly describe the complex WKB method for adiabatically perturbed
periodic Schr{\"o}dinger equations
\begin{equation}
  \label{G.2}
  -\frac{d^2}{dx^2}\psi(x)+(V(x)+W(\varepsilon
  x+\zeta))\psi(x)= E\psi(x), \quad x\in\R.
\end{equation}
Here, $V$ is $1$-periodic and $\varepsilon$ is a small positive
parameter; one assumes that $V\in L_{\rm loc}^{2}$ and that $W$ is
analytic in ${\mathcal D}(W)$, a neighborhood of the real line
($W$ is not necessarily periodic).
\smallpagebreak The parameter $\zeta$ is an auxiliary parameter
used to decouple the ``slow variable'' $\xi=\varepsilon x$ and the
``fast variable'' $x$. The idea of the method is to study
solutions of~\eqref{G.2} on the complex plane of $\zeta$ and, then
to recover information on their behavior in $x\in\R$. Therefore,
one studies solutions satisfying the  condition:
\begin{equation}
  \label{consistency:1}
  \psi(x+1,\zeta)=\psi (x,\zeta+\varepsilon)\quad\forall\zeta.
\end{equation}
On the complex plane of $\zeta$, there are certain domains on which
one can construct such solutions that, moreover, have simple
asymptotic behavior.
\subsection{Standard behavior of solutions}
\label{sec:stand-behav-solut}
We first define two analytic objects central to the complex WKB
method, the {\it complex momentum} and the {\it canonical Bloch
  solutions}. Then, we describe the {\it standard behavior} of the
solutions studied in the framework of the complex WKB method.
\subsubsection{The complex momentum}
\label{sec:kappa}
The complex momentum $\kappa$ is the main analytic object of the
complex WKB method. For $\zeta\in\mathcal{D}(W)$, the domain of
analyticity of the function $W$, it is defined by the formula
\begin{equation}
  \label{k->kappa}
  \kappa(\zeta)=k(\mathcal{E}(\zeta)),\quad
  \mathcal{E}(\zeta)=E-W(\zeta),
\end{equation}
Here, $k$ is the Bloch quasi-momentum of~\eqref{Ho}.
Relation~\eqref{k->kappa} ``translates'' properties of $k$ into
properties of $\kappa$. Hence, $\kappa$ is a multi-valued analytic
function, and that its branch points are related to the branch points
of the quasi-momentum by the relations
\begin{equation}
  \label{bp}
  E_l+W(\zeta)=E,\quad l=1,2,3\dots
\end{equation}
Let $\zeta_0$ be a branch point of $\kappa$. If $W'(\zeta_0)\ne
0$, then $\zeta_0$   is a branch point of square root type.
\smallpagebreak If $D\subset\mathcal{D}(W)$ is a simply connected set
containing no branch points of $\kappa$, we call it {\it regular}. Let
$\kappa_p$ be a branch of the complex momentum analytic in a regular
domain $D$. All the other branches analytic in $D$ are described by
the formulae:
\begin{equation}
  \label{allbr}
  \kappa_m^\pm  =\pm  \kappa_p+2\pi m,
\end{equation}
where $\pm $ and $m\in\Z$ are indexing the branches.
\subsubsection{Canonical Bloch solutions} 
\label{CBS}
To describe the asymptotic formulae of the complex WKB method, one
needs Bloch solutions to the equation
\begin{equation}
  \label{eq:52}
  -\frac{d^2}{dx^2}\psi(x)+V(x)\psi(x)={\mathcal E}(\zeta)\psi(x),\quad
  {\mathcal E}(\zeta)=E-W(\zeta),\quad x\in\R,
\end{equation}
that, moreover, are analytic in $\zeta$ on a given regular domain.
\smallpagebreak Pick $\zeta_0$ a regular point. Let
$\mathcal{E}_0=\mathcal{E}(\zeta_0)$. Assume that $\mathcal{E}_0\not
\in P\cup Q$. Let $U_0$ be a sufficiently small neighborhood of
${\mathcal E}_0$, and let $V_0$ be a neighborhood of $\zeta_0$ such
that ${\mathcal E}(V_0)\subset U_0$.  In $U_0$, we fix a branch of the
function $\sqrt{k'(\mathcal{E})}$ and consider
$\psi_\pm(x,\mathcal{E})$, two branches of the Bloch solution
$\psi(x,\mathcal{E})$, and $\Omega_\pm$, the corresponding branches of
$\Omega$. Put
\begin{equation}
  \label{canonicalBS}
  \Psi_\pm (x,\zeta)=
  q(\mathcal{E})\,e^{\int_{\mathcal{E}_0}^\mathcal{E} \Omega_\pm}
  \psi_\pm (x,\mathcal{E}),\quad q(\mathcal{E})=\sqrt{k'(\mathcal{E})},\quad
  \mathcal{E}=\mathcal{E}(\zeta).
\end{equation}
The functions $\Psi_\pm$ are called the {\it canonical Bloch solutions
  normalized at the point $\zeta_0$}.
\smallpagebreak The properties of the differential $\Omega$ imply
that the solutions $\Psi_\pm$ can be analytically continued from
$V_0$ to any regular domain $D$ containing $V_0$. \\
One has (see~\cite{Fe-Kl:03a})
\begin{equation}
  \label{Wcanonical}
  w(\Psi_+(\cdot ,\zeta),\Psi_-(\cdot ,\zeta))=w(\Psi_+(\cdot ,\zeta_0),\Psi_-(\cdot ,\zeta_0))=
  k'(\mathcal{E}_0)w(\psi_+(x,\mathcal{E}_0),\psi_-(x,\mathcal{E}_0))
\end{equation}
As $\mathcal{E}_0\not\in Q\cup\{E_l\}$, the Wronskian
$w(\Psi_+(\cdot ,\zeta),\Psi_-(\cdot ,\zeta))$ is non-zero.
\subsection{Solutions having standard asymptotic behavior}
\label{sec:standard-asymptotics}
Fix $E=E_0$. Let $D$ be a regular domain. Fix $\zeta_0\in D$ so that
$\mathcal{E}(\zeta_0)\not\in P\cup Q$. Let $\kappa$ be a branch of the
complex momentum continuous in $D$, and let $\Psi_\pm$ be the
canonical Bloch solutions defined on $D$, normalized at $\zeta_0$ and
indexed so that $\kappa$ be the quasi-momentum for $\Psi_+$.
\begin{Def}
  \label{def:1}
  Let $\sigma$ be either $+$ or $-$. We say that, in $D$, a solution
  $f$ has standard behavior (or standard asymptotics)
  $f\sim\exp(\sigma\frac{i}{\varepsilon} \int^{\zeta}
  \kappa\,d\zeta)\cdot\Psi_\sigma$ if
  \begin{itemize}
  \item there exists $V_0$, a complex neighborhood of $E_0$, and $X>0$
    such that $f$ is defined and satisfies~\eqref{G.2}
    and~\eqref{consistency:1} for any $(x,\zeta,E)\in [-X,X]\times
    D\times V_0$;
  \item $f$ is analytic in $\zeta\in D$ and in $E\in V_0$;
  \item for any $K$, a compact subset of $D$, there is $V\subset V_0$,
    a neighborhood of $E_0$, such that, for $(x,\zeta,E)\in
    [-X,X]\times K\times V$, $f$ has the uniform asymptotic
    \begin{equation}
      \label{stand:as}
      f=e^{\dsize\sigma\,\frac{i}{\varepsilon} \int^{\zeta}
        \kappa\,d\zeta}\, (\Psi_\sigma+o\,(1)),\quad \text{as}\quad
      \varepsilon\to 0,
    \end{equation}
  \item this asymptotic can be differentiated once in $x$ without
    loosing its uniformity properties.
\end{itemize}
\end{Def}
\subsection{Canonical domains}
\label{sec:Canonical-domains}
Canonical domains are important examples of domains where one can
construct solutions with standard asymptotic behavior. They are
defined using canonical lines.
\subsubsection{Canonical lines}
\label{sec:canonical-lines}
A curve is called {\it vertical} if it is connected, piecewise $C^1$,
and if it intersects the lines $\{\im\zeta=\Const\}$ at non-zero
angles $\theta$, \ $0<\theta<\pi$. Vertical curves are naturally
parameterized by $\im\zeta$.
\smallpagebreak Let $\gamma$ be a regular curve. On $\gamma$, fix
$\kappa$, a continuous branch of the complex momentum.
\begin{Def}
  \label{def:2}
  The curve $\gamma$ is {\it canonical} if it is vertical and such
  that, along $\gamma$,
  \begin{enumerate}
  \item $ \im\int^{\zeta}\kappa d\zeta$ is strictly monotonously
    increasing with $\im\zeta$,
  \item $\im\int^{\zeta} (\kappa-\pi)d\zeta$ is strictly monotonously
    decreasing with $\im\zeta$.
  \end{enumerate}
\end{Def}
\noindent Note that canonical lines are stable under small
$\mathcal{C}^{1}$-perturbations.
\subsubsection{Canonical domains}
\label{sec:canonical-domains}
Let $K$ be a regular domain. On $K$, fix a continuous branch of
the complex momentum, say $\kappa$. The domain $K$ is called {\it
canonical} if it is the union of curves canonical with respect to
$\kappa$ and connecting two points $\zeta_1$ and $\zeta_2$ located
on $\partial K$.
\smallpagebreak One has
\begin{Th}[\cite{Fe-Kl:03a}]
  \label{T5.1}
  Let $K$ be a bounded domain canonical with respect to $\kappa$.  For
  sufficiently small positive $\varepsilon$, there exists $(f_\pm )$,
  two solutions of~\eqref{G.2}, having the standard behavior in $K$:
  \begin{equation*}
    f_\pm \sim \exp\left(\pm \frac{i}{\varepsilon}
      \int_{\zeta_{0}}^{\zeta} \kappa d\zeta\right)\Psi_\pm   .
  \end{equation*}
  For any fixed $x\in\R$, the functions $f_\pm (x,\zeta)$ are analytic
  in $\zeta$ in the smallest strip $\{Y_1<\im\zeta<Y_2\}$ containing
  $K$.
\end{Th}
\noindent One easily calculates the Wronskian of the solutions
$f_\pm (x,\zeta)$ to get
\begin{equation}
  \label{W_of_f_pm}
  w(f_+,f_-)=w(\Psi_+,\Psi_-)+o(1).
\end{equation}
By~\eqref{Wcanonical}, for $\zeta$ in any fixed compact subset of $K$
and $\varepsilon$ sufficiently small, the solutions $f_\pm $ are
linearly independent.
\subsection{The strategy of the WKB method}
\label{strategy}
Our strategy to apply the complex WKB method is explained in great
detail in~\cite{Fe-Kl:03a}; we recall it briefly. First, we find a
canonical line. Roughly, we build it out of segments of some
``elementary'' curves described in section~\ref{pre-cl}. Then, we find
a ``local'' canonical domain $K$ ``enclosing'' this line, see
section~\ref{sec:LCD}. For this domain, we construct the solutions
$f_\pm$ using Theorem~\ref{T5.1}.
\smallpagebreak Second, we describe the asymptotic behavior of $f_\pm$
outside the domain $K$. Therefore, we use three general principles
presented in section~\ref{sec:m-t}.
\smallpagebreak Having investigated the behavior of $f_\pm$ for $-X\le
x\le X$ and a sufficiently large set of $\zeta$, we recover the
behavior of $f_\pm$ on the real line of $x$ by means of
condition~\eqref{consistency:1}.
\medpagebreak Below, we assume that $D$ is a regular domain, and
that $\kappa$ is a branch of the complex momentum analytic in $D$.
A {\it segment} of a curve is a connected, compact subset of that
curve.
\subsection{Local canonical domains}
\label{sec:LCD}
Let $\gamma\subset D$ be a canonical line (with respect to $\kappa$).
Denote its ends by $\zeta_{1}$ and $\zeta_{2}$.  Let a domain
$K\subset D$ be a canonical domain corresponding to the triple
$\kappa$, $\zeta_{1}$ and $\zeta_{2}$. If $\gamma\in K$, then, $K$ is
called a canonical domain {\it enclosing} $\gamma$.  One has
\begin{Le}[\cite{Fe-Kl:02}]
  \label{LCD}
  One can construct a canonical domain enclosing any given canonical
  curve.
\end{Le}
\noindent Canonical domains whose existence is a consequence of this
lemma are called {\it local}.
\subsection{Pre-canonical lines} 
\label{pre-cl}
To construct a local canonical domain we need a canonical line. To
construct such a line, we first build a pre-canonical line made of
some ``elementary'' curves.
\smallpagebreak Let $\gamma\subset D$ be a vertical curve. We call
$\gamma$ {\it pre-canonical} if it is a finite union of segments of
canonical lines and/or lines of Stokes type i.e. the level curves of
the harmonic functions $\zeta\mapsto\im\int^\zeta\kappa d\zeta$ or
$\zeta\mapsto\im\int^\zeta(\kappa-\pi)d\zeta$. One has
\begin{Pro}[\cite{Fe-Kl:02}]
  \label{pro:pcl:1}
  Let $\gamma$ be a pre-canonical curve. Denote the ends of $\gamma$
  by $\zeta_a$ and $\zeta_b$.\\
  For $V\subset D$, a neighborhood of $\gamma$ and $V_a\subset D$, a
  neighborhood of $\zeta_a$, there exists a canonical line
  $\gamma\subset V$ connecting the point $\zeta_b$ to a point in
  $V_a$.
\end{Pro}
\noindent When constructing pre-canonical lines, one uses lines of
Stokes type. To analyze them, one uses
\begin{Le}
  \label{le:s-l-1} 
  The lines of Stokes type of the family
  $\im\int^{\zeta}\kappa\,d\zeta=\Const$ are tangent to the vector
  field $\overline{\kappa(\zeta)}$; those of the family
  $\im\int^{\zeta}(\kappa-\pi)\,d\zeta=\Const$ are tangent to the
  vector field $\overline{\kappa(\zeta)}-\pi$.
\end{Le}
\subsection{Tools for computing global asymptotics}
\label{sec:m-t}
A set is said to be {\it constant} if it is independent of
$\varepsilon$.
\subsubsection{The Rectangle Lemma: asymptotics of increasing
  solutions}
\label{sec:rectangle-lemma}
The Rectangle Lemma roughly says that a solution $f$ preserve the
standard behavior along a line $\im\zeta=\Const$ as long as the
leading term of the standard asymptotics increases.
\smallpagebreak Fix $\eta_m<\eta_M$. Define $S=\{\zeta\in\C:\ 
\eta_m\le \im\zeta\le\eta_M\}$. Let $\gamma_1$ and $\gamma_2$ be two
vertical lines such that $\gamma_1\cap \gamma_2=\emptyset$.  Assume
that both lines intersect the strip $S$ at the lines $\im\zeta=\eta_m$
and $\im\zeta=\eta_M$, and that $\gamma_1$ is
situated to the left of $\gamma_2$.\\
Consider  the compact $R$  bounded by $\gamma_1$, $\gamma_2$ and
the boundaries of $S$. Let $D$=$R\setminus(\gamma_1\cup\gamma_2)$.
One has
\begin{Le}[The Rectangle Lemma~\cite{Fe-Kl:01b}]
  \label{Rectangle}
  Assume that the ``rectangle'' $R$ is contained in a regular domain.
  Let $f$ be a solution to~\eqref{G.2}
  satisfying~\eqref{consistency:1}. Then, for sufficiently small
  $\varepsilon$, one has
  \begin{description}
  \item[1] If $\im\kappa<0$ in $D$, and if $f$ has standard behavior
    $f\sim e^{\frac{i}{\varepsilon} \int_{\zeta_0}^{\zeta}\kappa
      d\zeta} \Psi_+$ in a neighborhood of $\gamma_1$, then, it has
    standard behavior in a constant domain containing the
    ``rectangle'' $R$.
  \item[2] If $\im\kappa>0$ in $D$, and if $f$ has standard behavior
    $f\sim e^{\frac{i}{\varepsilon} \int_{\zeta_0}^{\zeta}\kappa
      d\zeta} \Psi_+$ in a neighborhood of $\gamma_2$, then, it has
    the standard behavior in a constant domain containing the
    ``rectangle'' $R$.
  \end{description}
\end{Le}
\noindent Lemma~\ref{Rectangle} was proved in~\cite{Fe-Kl:01b}
where one can find more details and references.
\subsubsection{The Adjacent Canonical Domain Principle}
\label{sec:addCD}
This principle complements the Rectangle Lemma; it allows us to obtain
the asymptotics of decreasing solutions.
\smallpagebreak Let $\gamma$ be a vertical curve.  Let $S$ be the
minimal strip of the form $\{C_1\le \im\zeta\le C_2\}$ containing
$\gamma$. Let $U\subset S$ be a regular domain such that
$\gamma\subset \partial U$. We say that $U$ is adjacent to $\gamma$.
One has
\begin{Pro}[The Adjacent Canonical Domain Principle~\cite{Fe-Kl:01b}]
  \label{AddCD}
  Let $\gamma$ be a canonical line. Assume that $f$, a solution
  to~\eqref{G.2} satisfying~\eqref{consistency:1} has standard
  behavior in a domain adjacent to $\gamma$. Then, $f$ has the
  standard behavior in any bounded canonical domain enclosing
  $\gamma$.
\end{Pro}
\subsubsection{Adjacent canonical domains}
\label{sec:adjac-canon-doma}
To apply the Adjacent Canonical Domain Principle, one needs to describe
canonical domains enclosing a given canonical line. It can be quite
difficult to find a ``maximal'' canonical domain enclosing a given
canonical line. In practice, one uses ``simple'' canonical domains
described in 
\begin{Le}[The Trapezium Lemma~\cite{Fe-Kl:03a}]
  \label{trapezium-le}
  Let $\gamma_0$ be a canonical line. 
                                %
  \begin{floatingfigure}{5cm}
    \begin{center}
      \includegraphics[bbllx=60,bblly=662,bburx=219,bbury=722,width=5cm]{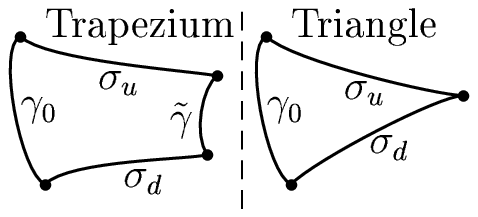}
    \end{center}
  \end{floatingfigure}
                                %
  \noindent Let $U$ be a domain adjacent to $\gamma$, a canonical line
  containing $\gamma_0$ as an internal segment. Assume that, in $U$,
  $\im\kappa\ne 0$. Let $\sigma_u$ (resp. $\sigma_d$) be the line of
  Stokes type beginning at the upper (resp.  lower) end of $\gamma_0$
  and going in $U$ downward (resp.  upward) from this point.\\
  One has:
  \begin{description}
  \item[Trapezium case] Let $T\subset U$ be a regular domain bounded
    by $\sigma_u$, $\sigma_d$, $\gamma_0$ and $\tilde\gamma$, one more
    canonical line not intersecting $\gamma_0$. Then, $T$ is part of a
    canonical domain enclosing $\gamma_0$.
  \item[Triangle case] Assume that $\sigma_u$ intersects $\sigma_d$.
    Let $T\subset U$ be a regular domain bounded by $\sigma_u$,
    $\sigma_d$ and line $\gamma_0$. Then, $T$ is part of a canonical
    domain enclosing $\gamma_0$.
  \end{description}
\end{Le}
\vskip.1cm\noindent There always exists a canonical line containing
$\gamma_0$; moreover, if $\im\kappa\ne0$ in $U$, then, the lines
$\sigma_d$ and $\sigma_u$ described in the Trapezium Lemma always
exist (see Lemma 5.3 from~\cite{Fe-Kl:03a}).
\subsubsection{The Stokes Lemma}
\label{sec:St-Lemme}
The domains where one justifies the standard behavior using the
Adjacent Canonical Domain Principle are often bounded by Stokes lines
(see definition below) beginning at branch points of the complex
momentum.  The Stokes Lemma allows us to justify the standard behavior
beyond these lines by ``going around'' the branch points.
\smallpagebreak {\it Notations and assumptions.\/} Assume that
$\zeta_0$ is a branch point of the complex momentum such that
$W'(\zeta_0)\ne 0$.
\begin{Def}
  \label{def:3}
  The Stokes lines starting at $\zeta_0$ are the\\ curves $\gamma$
  defined by
  \begin{equation*}
    \hskip-6cm
    \im\int_{\zeta_0}^\zeta(\kappa(\zeta)-\kappa(\zeta_0))d\zeta=0,\quad
    \zeta\in\gamma.
  \end{equation*}
\end{Def}
\noindent The angles between the Stokes lines at $\zeta_0$ are equal to
$2\pi/3$. We denote them by $\sigma_1$, $\sigma_2$ and $\sigma_3$ so
that $\sigma_1$ is vertical at $\zeta_0$ (see
Fig.~\ref{stokes:fig:1}).
\smallpagebreak Let $\tilde\sigma_1$ be a (compact) segment of
$\sigma_1$ which begins at $\zeta_0$, is vertical and contains only
one branch point, i.e. the point $\zeta_0$.
\vskip.1cm\noindent Let $V$ be a neighborhood of $\tilde\sigma_1$.
Assume that $V$ is so small that the Stokes lines $\sigma_1$,
$\sigma_2$ and $\sigma_3$ divide it into three sectors. We denote them
by $S_1$, $S_2$ and $S_3$ so that $S_1$ be situated between $\sigma_1$
and $\sigma_2$, and the sector $S_2$ be between $\sigma_2$ and
$\sigma_3$ (see Fig.~\ref{stokes:fig:1}).
\smallpagebreak{\it The statement.\/} In~\cite{Fe-Kl:03a}, we have
proved
\begin{Le}[Stokes Lemma]
  \label{st-lm}
  Let $V$ be sufficiently small. 
                                %
  \begin{floatingfigure}{7cm}
    \begin{center}
      \includegraphics[bbllx=71,bblly=577,bburx=360,bbury=721,width=7cm]{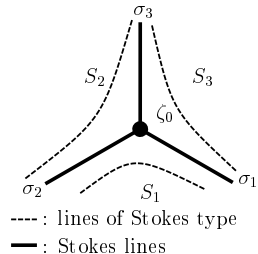}
    \end{center}
    \caption{The Stokes lines in a neighborhood of a branch point}
    \label{stokes:fig:1}
  \end{floatingfigure}
                                %
                                %
  Let $f$ be a solution to~\eqref{G.2}
  satisfying~\eqref{consistency:1} that has standard behavior $f\sim
  e^{\frac i\varepsilon\int^\zeta\kappa d\zeta}\Psi_+$ inside the
  sector $S_1\cup \sigma_2\cup S_2$ of $V$. Moreover, assume that, in
  $S_1$ near $\sigma_1$, one has $\im\kappa(\zeta)>0$ if $S_1$ is to
  the left of $\sigma_1$ and $\im\kappa(\zeta)>0$ otherwise.\\
  Then, $f$ has standard behavior inside $V\setminus \sigma_1$, the
  leading term of the asymptotics being obtained by analytic
  continuation from $S_1\cup\sigma_2\cup S_2$ into $V'$.
\end{Le}
Comments and details on this lemma can be found in~\cite{Fe-Kl:03a}.


%
\section{A consistent solution}
\label{sec:f}
\noindent We now begin the construction of the consistent basis the
monodromy matrix of which we compute. Recall that $V$ and $W$ satisfy
the hypothesis (H), and fix $E=E_0\in J$, an interval satisfying the
hypothesis (A1) -- (A3),
\smallpagebreak In the present section, by means of the complex WKB
method, we construct and study $f$ a solution of~\eqref{family}
satisfying the consistency condition~\eqref{consistency}.
\smallpagebreak To use the complex WKB method, we
rewrite~\eqref{family} in terms of the variables
\begin{equation}\label{new-variables}
x:=x-z,\quad\text{and}\quad  \zeta=\varepsilon z.
\end{equation}
It then takes the form~\eqref{G.2}. In the new variables, the
consistency condition~\eqref{consistency}
becomes~\eqref{consistency:1}
\smallpagebreak We first describe the complex momentum and Stokes
lines. Then, we construct a local canonical domain, hence, a
consistent solution to~\eqref{G.2} by Theorem~\ref{T5.1}. Finally,
using the continuation tools, we describe global asymptotics of this
solution.
\subsection{The complex momentum}
\label{sec:complex-momentum}
We begin with the analysis of the mapping $\mathcal{E}: \zeta\to
E-W(\zeta)$.
\subsubsection{The set $W^{-1}(\R)$}
\label{sec:set-w-1r}
As $E\in\R$, $\mathcal{E}^{-1}(\R)$ coincides with $W^{-1}(\R)$.
\smallpagebreak The set $W^{-1}(\R)$ is $2\pi$-periodic. It consists
of the real line and of complex branches (curves) symmetric with
respect to the real line. There are complex branches separated from
the real line, and complex branches beginning at the real extrema of
$W$. These do not return to the real line.
\smallpagebreak Consider an extremum of $W$ on the real line, say
$\zeta_0$. By assumption (H), it is non-degenerate. This implies that,
near $\zeta_0$, the set $W^{-1}(\R)$ consists of a real segment, and
of a ``complex'' curve symmetric with respect to the real axis,
intersecting the real axis at $\zeta_0$ only. This curve is orthogonal
to the real line at $\zeta_0$.
\smallpagebreak For $Y>0$, we let $\mathcal{S}_Y=\{-Y\le\im\zeta\le
Y\}$. We assume that $Y$ is so small that
\begin{itemize}
\item $S_Y$ is contained in the domain of analyticity of $W$;
\item the set $W^{-1}(\R)\cap {\mathcal S}_Y$ consists of the real
  line and of the complex lines passing through the real extrema of
  $W$;
\end{itemize}
\noindent For such $Y$, and if $m=0$, the set
$W^{-1}(\R)\cap{\mathcal S}_Y$ is shown in Fig.~\ref{fig:ibm12}.
\subsubsection{Branch points} 
\label{sec:branch-points}
The branch points of the complex momentum are related to the branch
points of the Bloch quasi-momentum by equation~\eqref{bp}. So, they
lie on $W^{-1}(\R)$, and form a $2\pi$-periodic set.
\smallpagebreak Consider the branch points situated in the interval
$[0,2\pi)$ of the real line. Recall that, by assumption, the function
$\zeta\mapsto W(\zeta)$ has two extrema in $[0,2\pi]$: a maximum at
$\zeta=0$ and a minimum at $\zeta^*$, $0<\zeta^*<2\pi$.  The mapping
$\mathcal{E}$ is monotonous on each of the intervals $I_-=[0,\zeta_*]$
and $I_+=[\zeta^*,2\pi]$ and maps both $I_\pm$ onto the interval
$[E-W_+,E-W_-]$. Under the hypotheses (A1) -- (A3), the interval
$[E-W_+,E-W_-]$ contains $2m+2$ branch points of the Bloch
quasi-momentum, namely the points $E_j$ for $j=2n-1,2n\dots 2(n+m)$.
Therefore, on $[0,2\pi]$, one has $4m+4$ branch points of the complex
momentum $\zeta_j^\pm$, \ $ j=2n-1,2n,\dots 2(n+m)$, such that
$\zeta_j^\pm\in I_\pm$ and $\mathcal{E}(\zeta_j^\pm)=E_j$. They
satisfy the inequalities
\begin{equation}
  \label{bp:seq}
  0<\zeta_{2n-1}^-<\zeta_{2n}^-<\dots<\zeta_{2(n+m)}^-<\zeta^\ast<
  \zeta_{2(n+m)}^+<\zeta_{2(n+m)-1}^+<\dots<\zeta_{2n-1}^+<2\pi.
\end{equation}
\smallpagebreak The complex branch points all lie on $W^{-1}(\R)$.
Reducing $Y$, at no loss of generality, we from now on assume that the
strip ${\mathcal S}_Y$ contains only the real branch points of the
complex momentum.
\subsubsection{The sets $Z$ and $G$}
\label{sec:ZandG}
Consider the gaps and bands of the periodic operator~\eqref{Ho}.  Let
$Z$ and $G$ respectively be the pre-image (with respect to
$\mathcal{E}$) of the bands and the pre-image of the gaps. Clearly,
$Z\cup G= W^{-1}(\R)$, and $Z\cap G=\emptyset$. Connected components
of $Z$ and $G$ are separated by branch points of the complex momentum.
%
%
\begin{floatingfigure}{4cm}
  \begin{center}
    \includegraphics[bbllx=66,bblly=634,bburx=190,bbury=721,width=4cm]{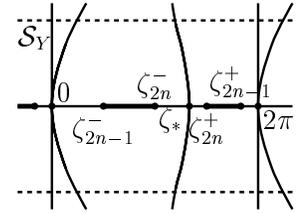}
  \end{center}
  \caption{The branch points when $m=0$}\label{fig:ibm12}
\end{floatingfigure}
%
%
\smallpagebreak The set $Z$ is $2\pi$-periodic. Consider the part of
$Z$ situated in the interval $[0,2\pi]$. Consider $\mathcal Z$, the
collection of subintervals of $[0,2\pi]$ defined in
section~\ref{res:real-branches}. One has
\begin{equation}
  \label{pre-bands}
  \hskip-3cm  
  \mathfrak{z}_j^-=[\zeta_{2j-1}^-, \zeta_{2j}^-],\quad
  \mathfrak{z}_j^+=[\zeta_{2j}^+,\zeta_{2j-1}^+],\quad j=n,n+1,\dots n+m.
\end{equation}
Under our assumptions on $Y$, the set $Z\cap S_Y$ consists of the
intervals of $\mathcal Z$ and their $2\pi$-translates.
\smallpagebreak Define the finite collection of intervals $\mathcal G$
as in section~\ref{res:cl-cur}.
One has
\begin{equation}
  \label{pre-gaps}
  \begin{split}
    \hskip-3cm
    \mathfrak{g}_j^-=(\zeta_{2j}^-, \zeta_{2j+1}^-),&\quad
    \mathfrak{g}_j^+=(\zeta_{2j+1}^+,\zeta_{2j}^+),\quad
    j=n,n+1,\dots  n+m-1,\\
    \hskip-3cm
    \mathfrak{g}_{n-1}&=(\zeta_{2n-1}^+-2\pi,\zeta_{2n-1}^-),\quad
    \mathfrak{g}_{n+m}=(\zeta_{2(n+m)}^-,\zeta_{2(n+m)}^+),
  \end{split}
\end{equation}
The set $G\cap S_Y$ consists of the following connected components:
\begin{itemize}
\item the intervals $\mathfrak{g}_j^\pm$, $j=n,\dots n+m-1$;
\item the connected component containing $\zeta=0$ (it is the union of
  the interval $\mathfrak{g}_{n-1}$ and the complex branch of
  $W^{-1}(\R)\cap S_{Y}$ passing through $0$);
\item the connected component containing $\zeta=\zeta^*$ (it is the
  union of the interval $\mathfrak{g}_{n+m}$ and the complex branch of
  $W^{-1}(\R)\cap S_{Y}$ passing through $\zeta^*$);
\item all the $2\pi$-translates of the curves mentioned above.
\end{itemize}
\subsubsection{The main branch of the complex momentum}
\label{sec:complex-momentum-1}
Introduce {\it the main branch} of the complex momentum. In the strip
$\{0<\im\zeta<Y\}$, consider the domain $D_p$ between the complex
branches of $W^{-1}(\R)$ beginning at $0$ and at $\zeta^*$. It is
regular and ${\mathcal E}=E-W(\zeta)$ conformally maps it onto a
domain in the upper half of the complex plane. We define the main
branch of $\kappa$ by the formula
\begin{equation}
  \label{kappaP}
  \kappa_p(\zeta)=k_p(E-W(\zeta)),\quad \zeta\in D_p,
\end{equation}
where $k_p$ is the main branch of the Bloch quasi-momentum of the
periodic operator~\eqref{Ho} (see section~\ref{SS3.2}). Clearly, $\im
\kappa_p>0$ in $D_p$ and the function $\kappa_p$ is continuous up to
the boundary of $D_p$. Its behavior at the boundary of $D_p$ reflects
the behavior of $k_p$ along the real line. In particular, for each
$j=n,n+1\dots,n+m$, it is monotonously increasing on
$\mathfrak{z}_j^-$ and maps it onto $[\pi(j-1),\pi j]$.
\subsubsection{The Stokes lines}
\label{sec:stokes-lines-1}
Consider the Stokes lines beginning at $\zeta_{2n-1}^-$ and
$\zeta_{2n}^-$. As $\kappa_p$ is real on $\mathfrak{z}_n^-=
[\zeta_{2n-1}^-, \zeta_{2n}^-]$, this interval is a Stokes line both
for $\zeta_{2n-1}^-$ and $\zeta_{2n}^-$.  Pick one of these points. As
$W'\ne 0$ at this point, the angles between the Stokes lines beginning
at it are equal to $2\pi/3$. So, one of the Stokes lines is going
upwards, one is going downwards.  These two Stokes lines are symmetric
with respect to the real line.
\smallpagebreak Denote by $\sigma_1$  the Stokes line starting
from $\zeta_{2n-1}^-$ going downwards, and denote by $\sigma_2$ be
the Stokes line beginning at $\zeta_{2n}^-$ and going upwards (see
Fig.~\ref{fig:ibm8}).
\smallpagebreak The lines $\sigma_1$ and $\sigma_2$ are vertical in
$S_Y$. Indeed, a Stokes line stays vertical as long as $\im\kappa\ne
0$; the imaginary part of the complex momentum vanishes only on $Z$,
and $Z\cap S_Y\subset \R$.
\smallpagebreak Reducing $Y$ if necessary, we can assume that
$\sigma_1$ and $\sigma_2$ intersect the boundaries of $S_Y$.

\subsection{Local construction of the solution $f$}
\label{sec:local-constr-solut}
We construct $f$ on a local canonical domain. To construct a local
canonical domain, we need a canonical line. To find a canonical line,
we first build a pre-canonical line.
\subsubsection{Pre-canonical line}
\label{sec:pre-canonical-line}
\begin{figure}[h]
  \centering
  \includegraphics[bbllx=71,bblly=513,bburx=601,bbury=723,height=5cm]{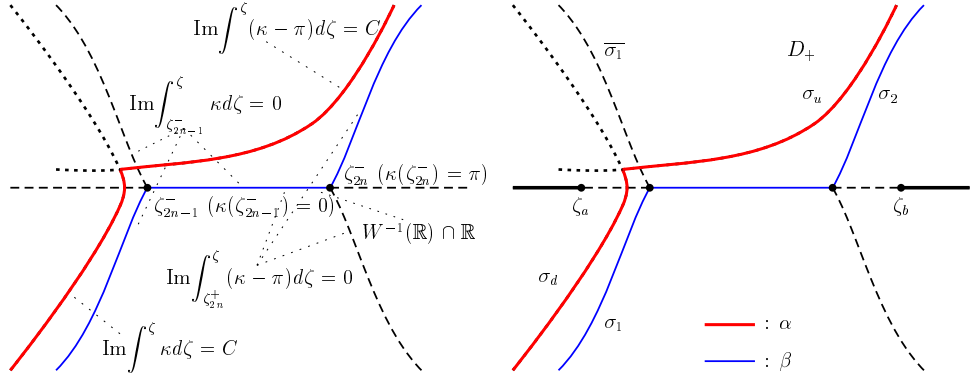}
  \caption{The construction of the pre-canonical curve}\label{fig:ibm14}
\end{figure}

Consider the curve $\beta$ which is the union of the Stokes lines
$\sigma_1$, $[\zeta_{2n-1}^-, \zeta_{2n}^-]$ and $\sigma_2$. Let us
construct $\alpha$, a pre-canonical line close to the line $\beta$. It
goes around the branch points of the complex momentum as shown
in~Fig.~\ref{fig:ibm14}.\\
When speaking of $\kappa_p$ along $\alpha$, we mean the branch of the
complex momentum obtained of $\kappa_p$ by analytic continuation along
this line (the analytic continuation can be done using
formula~\eqref{k->kappa}).
\smallpagebreak Actually, the line $\alpha$ will be pre-canonical with
respect to the branch of the complex momentum related to $\kappa_p$ by
the formula:
\begin{equation}
  \label{kappaPkappa}
  \kappa=\left\{\begin{array}{ll}
      \kappa_p-\pi (n-1) & \quad \text{if \ } n \text{\ is \ odd}\\
      \pi n -\kappa_p & \quad \text{if \ } n \text{\ is \ even}
    \end{array}\right.
  \quad\quad \zeta\in D_p.
\end{equation}
In view of~\eqref{allbr}, $\kappa$ is indeed a branch of the complex
momentum. We prove
\begin{Le}
  \label{pro:2}
  Fix $\delta>0$. In the $\delta$-neighborhood of $\beta$, to the left
  of $\beta$, there exists $\alpha$, a line pre-canonical with respect
  to the branch $\kappa$ such that at its upper end $\im\zeta>Y$, and
  at its lower end $\im \zeta<-Y$.
\end{Le}
\demo We consider only the case $n$ odd. The analysis of
the case of $n$ even is similar. Note that, for $n$ odd, 
formula~\eqref{kappaPkappa} implies that $\kappa(\zeta_{2n-1}^-)=0$, and
$\kappa(\zeta_{2n}^-)=\pi$. So, the Stokes lines $\sigma_1$ and
$\sigma_2$ satisfy the equations
$\im\int_{\zeta_{2n-1}^-}^\zeta\kappa\,d\zeta=0$ and
$\im\int_{\zeta_{2n}^-}^\zeta(\kappa-\pi)\,d\zeta=0$.
\smallpagebreak The pre-canonical line is constructed of $l_u$ and
$l_d$, two "elementary lines" that are segments of $\sigma_u$ and
$\sigma_d$, two lines of Stokes type shown in Fig.~\ref{fig:ibm14}.
Let us describe them more precisely.\\
%
%
\begin{floatingfigure}{4.5cm}
  \begin{center}
    \includegraphics[bbllx=71,bblly=606,bburx=219,bbury=721,width=4.5cm]{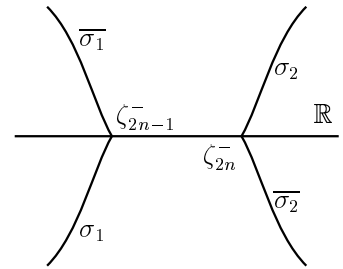}
  \end{center}
  \caption{Stokes lines}\label{fig:ibm8}
\end{floatingfigure}
%
Pick $\zeta_u$, a point of the line $\im\zeta=Y$, to the left of
$\beta$ close enough to it.  The line $\sigma_u$ is the line of Stokes
type $\im\int_{\zeta_u}^\zeta(\kappa-\pi) d\zeta=\Const$ containing
$\zeta_u$.  Note that all three curves, $\sigma_u$, $\sigma_2$ and
$[\zeta_{2n-1}^-,\zeta_{2n}^-]\subset\R$, belong to the same family of
curves $\im\int^\zeta(\kappa-\pi) d\zeta=\Const$. This implies that,
if $\zeta_u$ is close enough to $\sigma_2$, then, below $\zeta_u$,
$\sigma_u$ goes arbitrarily close to $\sigma_2\cup
[\zeta_{2n-1}^-,\zeta_{2n}^-]$ staying to the left of $\sigma_2$
and above $[\zeta_{2n-1}^-,\zeta_{2n}^-]$. We omit elementary details.\\
Pick $\zeta_d$, a point of the line $\im\zeta=-Y$, to the left of
$\beta$ close enough to it.  The line $\sigma_d$ is a line of Stokes
type $\im\int_{\zeta_d}^\zeta(\kappa-\pi) d\zeta=\Const$ containing
$\zeta_d$.  Note that all the three curves $\sigma_d$, $\sigma_1$ and
$\overline{\sigma_1}$ belong to the same family of curves
$\im\int^\zeta \kappa d\zeta=\Const$.  This implies that, if $\zeta_d$
is close enough to $\sigma_1$, then, above $\zeta_d$, $\sigma_d$ goes
arbitrarily close to the line $\sigma_1\cup \overline{\sigma_1}$ and
stays to the left of it. We omit the details.\\
Note that, by Lemma~\ref{le:s-l-1}, the curves $\sigma_u$ and
$\sigma_d$ are tangent to the vector fields $\overline{\kappa}-\pi$
and $\overline{\kappa}$ respectively.  This implies in particular that
each of the curves $\sigma_u$ and $\sigma_d$ stays vertical as long as
it does not intersect the set $Z$.\\
If $\sigma_u$ and $\sigma_d$ are chosen close enough to $\beta$, they
intersect one another in a neighborhood of $\zeta_{2n-1}^-$.  Indeed,
consider first $\sigma_u$. It goes from $\zeta_u$ downwards staying to
the left of $\sigma_2$ and above $[\zeta_{2n_1}^-,\zeta_{2n}^-]$.  So,
staying vertical, it has to intersect the Stokes line
$\overline{\sigma_1}$ above $\zeta_{2n-1}^-$, the beginning of
$\overline{\sigma_1}$. The intersection is transversal (as, first,
$\overline{\sigma_1}$ is tangent to the vector field
$\overline{\kappa}$, second, $\sigma_u$ is tangent to the vector field
$\overline{\kappa}-\pi$, and, third, at the intersection point,
$\im\kappa\ne 0$). If $\sigma_d$ is sufficiently close to
$\overline{\sigma_1}$, then $\sigma_u$ also intersects $\sigma_d$
transversally.
\vskip.1cm\noindent We choose the intersection point as the lower end
of $l_u$ and the upper end of $l_d$.  As $\sigma_u$ and $\sigma_d$ are
defined and are vertical somewhat outside $S_Y$, we can assume that
the upper end of $l_u$ is above the line $\im\zeta=Y$, that the lower
end of $l_d$ is below the line $\im\zeta=-Y$, and that both $l_d$ and
$l_u$ are vertical.\\
The lines $l_u$ and $l_d$ then form a pre-canonical line $\alpha$; it
can be chosen as close to the line $\beta$ as desired and, in
particular, inside the $\delta$-neighborhood of $\beta$.  This
completes the proof of Lemma~\ref{pro:2}.\qed
%
\subsubsection{Local canonical domain and a solution $f$}
\label{sec:local-canon-doma-1}

By Proposition~\ref{pro:pcl:1}, as close to $\alpha$ as desired, one
can find a canonical line $\gamma$. We construct $\gamma$ in the left
part of the $\delta$-neighborhood of $\beta$.
\smallpagebreak By Lemma~\ref{LCD}, there is $K$ a canonical domain
enclosing $\gamma$. We can assume that it is situated in arbitrarily
small neighborhood of $\gamma$. So, we construct $K$ in (the left part
of) the $\delta$-neighborhood of $\beta$.
\smallpagebreak By Theorem~\ref{T5.1}, on the canonical domain $K$, we
construct $f$, a solution to~\eqref{G.2} that has standard behavior
$f\sim \exp\left(\frac i\varepsilon\int^\zeta \kappa
  d\zeta\right)\,\Psi_+$ in $K$. We fix the normalization of this
solution later.
\subsection{Asymptotics of $f$ outside $K$}
\label{sec:asympt-f-outs}
Recall that $f$ is analytic in $\{Y_1<\im\zeta<Y_2\}$, the minimal
strip containing $K$.
\subsubsection{The results}
\label{sec:results-1}
Denote by $\zeta_a$ and $\zeta_b$ the branch points situated on $\R$
and, respectively, closest to $\zeta_{2n-1}^-$ on its left and closest
to $\zeta_{2n}^-$ on its right. Let $\mathcal D$ be a regular domain
obtained by cutting $S_Y$ along the Stokes lines $\sigma_1$, and
$\sigma_2$ and along the real intervals $(-\infty,\zeta_a]$ and
$[\zeta_b,\infty)$, see Fig.~\ref{fig:ibm15}. We prove
\begin{Pro}
  \label{pro:f:global} 
  If $\delta$ is sufficiently small, then, $f$ has standard behavior
  \begin{equation}
    \label{f:as}
    f=e^{\frac i\varepsilon\int_{\zeta_0}^\zeta\kappa
      d\zeta}\left(\Psi_+(x,\zeta,\zeta_0)+o(1)\right)
  \end{equation}
  in the whole domain $\mathcal D$.
\end{Pro}
\noindent The rest of this section is devoted to the proof of
Proposition~\ref{pro:f:global}. The proof is naturally divided into
``elementary'' steps. At each step, applying just one of the three
continuation tools (i.e. the Rectangle Lemma, the Adjacent domain
principle and the Stokes Lemma), we justify the standard behavior of
$f$ on one more subdomain of $\mathcal{D}$. Fig.~\ref{fig:ibm15} shows
where we use each of the continuation principles. Full straight arrows
indicate the use of the Rectangle Lemma, circular arrows indicate the
use of the Stokes Lemma, and, in the hatched domains, we use the
Adjacent Canonical Domain Principle.
\smallpagebreak Again, the analysis of the cases $n$ odd and $n$ even
are analogous. For sake of definiteness, we assume that $n$ odd and
consider only this case.  We shall use
\begin{Le}
  \label{im-kappa:sign}
  If $n$ is odd, then, in $\mathcal D\setminus
  [\zeta_{2n-1}^-,\zeta_{2n}^-]$,
  \begin{itemize}
  \item to the left of the Stokes lines $\sigma_1$ and $\sigma_2$, one
    has $\im\kappa>0$;
  \item to the right of these Stokes lines, one has $\im\kappa<0$.
  \end{itemize}
\end{Le}
\demo The sign of $\im\kappa$ remains the same in any regular domain
not intersecting $Z$. Moreover, the sign of $\im\kappa$ changes to the
opposite one as $\zeta$ intersects a connected component of $Z$ at a
point where $W'\ne 0$. So, in the connected subdomain of
$\mathcal{D}\setminus[\zeta_{2n-1}^-,\zeta_{2n}^-]$ to the left of
$\sigma_1\cup[\zeta_{2n-1}^-,\zeta_{2n}^-]\cup\sigma_2$, one has
$\im\kappa=\im(\kappa_p-\pi(n-1))=\im\kappa_p>0$ (here, we have
used~\eqref{kappaPkappa} for $n$ odd). To come from this subdomain to
the connected subdomain of $\mathcal{D}\setminus
[\zeta_{2n-1}^-,\zeta_{2n}^-]$ to the right of
$\sigma_1\cup[\zeta_{2n-1}^-,\zeta_{2n}^-]\cup\sigma_2$, one has to
intersect the interval $[\zeta_{2n-1}^-,\zeta_{2n}^-]$ which is a
connected component of $Z$.  So, in the right subdomain, one has
$\im\kappa<0$. \qed
\subsubsection{Behavior of $f$ between the lines $\gamma$ and $\beta$}
\label{sec:behavior-f-between}
To justify the standard asymptotics of $f$ in $\mathcal D$ between the
lines $\gamma$ and $\beta$, we use the Adjacent Canonical Domain
Principle.  Therefore, we need to describe a canonical domain
enclosing $\gamma$ (more precisely, the part situated between $\gamma$
and $\beta$). We do this by means of the Trapezium
Lemma~\ref{trapezium-le} (first statement).
%
%
\begin{floatingfigure}{6cm}
  \centering
  \includegraphics[bbllx=71,bblly=606,bburx=233,bbury=721,height=4cm]{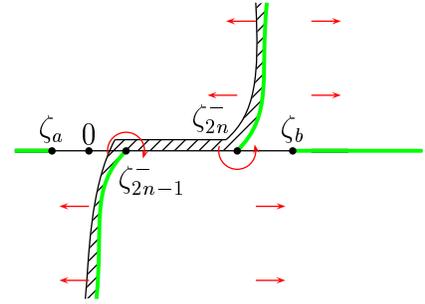}
  \caption{How to ``continue'' the asymptotics of $f$}\label{fig:ibm15}
\end{floatingfigure}
%
%
Let us describe the domain $U$ and the curves $\gamma_0$, $\sigma_d$
and $\sigma_u$ needed to apply the Trapezium Lemma~\ref{trapezium-le}.
\smallpagebreak {\it The domain $U$.\/} It is the domain bounded by
$\beta$, $\gamma$ and the lines $\im\zeta=\Const$ containing the ends
of $\gamma$.
\smallpagebreak In view of Lemma~\ref{im-kappa:sign}, choosing the
ends of $\gamma$ closer to the lines $\im\zeta=\pm Y$ if necessary, we
can assume that $\im\kappa>0$ in the domain $U$.
\smallpagebreak {\it The line $\sigma_u$.\/} As the line $\sigma_u$,
we take the line which belongs to the family $\im\int^{\zeta}\kappa
d\zeta=\Const$ and intersects $\beta$ at $\tilde\zeta_u$ satisfying
$\im\tilde\zeta_u=Y$. Recall that $\gamma$ is constructed in the
$\delta$-neighborhood of $\beta$ where $\delta$ can be fixed
arbitrarily small. One has
\begin{Le}
  \label{le:1}
  The line $\sigma_u$ enters $U$ at $\tilde\zeta_u$ and goes upwards.
  If $\delta$ is sufficiently small, then, $\sigma_u$ intersects
  $\gamma$ at an internal point of $\gamma$.
\end{Le}
\demo The main tool of this proof is Lemma~\ref{le:s-l-1}. Below, we
use it without further notice. Recall that, above the real line,
$\beta$ coincides with the Stokes line $\sigma_2$. So, it is tangent
to the vector field $\overline{\kappa(\zeta)-\pi}$. The line
$\sigma_d$ is tangent to the vector field $\overline{\kappa(\zeta)}$.
In $U$, and in particular at $\tilde\zeta_u$, one has $\im\kappa>0$.
Therefore, at $\tilde\zeta_u$, the tangent vector to $\beta$ (oriented
upwards) is directed to the right with respect to the tangent vector
to $\sigma_u$ (oriented upwards). So, $\sigma_u$ enters $U$ at
$\tilde\zeta_u$ going upwards.  As $\im\kappa\ne 0$ in $U$, it stays
vertical in $U$. Note that $\sigma_u$ is independent of $\delta$. So,
if $\delta$ is small enough, then $\sigma_u$ intersects $\gamma$. \qed
\smallpagebreak {\it The line $\sigma_d$.\/} It is the line of Stokes
type $\im\int^{\zeta}\kappa d\zeta=\Const$ that intersects $\beta$ at
$\tilde\zeta_d$ satisfying $\im\tilde\zeta_d=-Y$. One has
\begin{Le}
  \label{le:2}
  The line $\sigma_d$ enters $U$ at $\tilde\zeta_d$ and goes
  downwards. If $\delta$ is sufficiently small, then $\sigma_d$
  intersects $\gamma$ at an internal point.
\end{Le}
\noindent The proof of this lemma being similar to the proof of
Lemma~\ref{le:1}, we omit it.
\smallpagebreak{\it The line $\gamma_0$.\/} We choose $\delta$ so that
both $\sigma_u$ and $\sigma_d$ intersect $\gamma$. Then, $\gamma_0$ is
the segment of $\gamma$ between its intersection points with
$\sigma_d$ and $\sigma_u$.
\subsubsection{Describing the curve $\tilde\gamma$}  
\label{sec:descr-curve-tild}
Let us describe the canonical line $\tilde\gamma$ needed to apply the
first variant of the Trapezium Lemma. As for $\gamma$, using
Lemma~\ref{pro:2}, we can construct $\tilde\gamma$ so that it be
arbitrarily close to $\beta$ and strictly between $\gamma_0$ and
$\beta$.
\smallpagebreak As $\sigma_u$ and $\sigma_d$ intersect $\gamma$ and
$\beta$, they also intersect $\tilde \gamma$.
\subsubsection{Completing the analysis}
\label{sec:adp:1}
By the Trapezium Lemma, the domain bounded by $\gamma_0$, $\sigma_u$,
$\sigma_d$ and $\tilde\gamma$ is a part of the canonical domain
enclosing $\gamma_0$. So, by the Adjacent Canonical Domain Principle,
$f$ has the standard behavior here.
\smallpagebreak As $\tilde \gamma$ can be constructed arbitrarily
close to $\beta$, we conclude that $f$ has the standard behavior in
the whole domain bounded by $\beta$, $\gamma$ and the lines
$|\im\zeta|=Y$.
\subsection{Behavior of $f$ to the left of $\gamma$}
\label{ex:R}
We justify the standard behavior of $f$ in $\mathcal{D}$ to the left
of $\gamma$ by means of the Rectangle Lemma.
\smallpagebreak Let us describe $R$, the rectangle used to apply
Lemma~\ref{Rectangle}: it is the part of $\mathcal{D}$ between the
canonical line $\gamma_1=\gamma$ and a vertical line in $\mathcal{D}$,
say $\gamma_2$, staying to the left of $\gamma$ and going from
$\im\zeta=-Y$ to $\im\zeta=Y$.
\smallpagebreak By Lemma~\ref{im-kappa:sign}, in $R$, the imaginary
part of $\kappa$ is positive. Moreover, $f$ has standard behavior in a
neighborhood of $\gamma$. So, by the Rectangle Lemma, $f$ has standard
behavior in $R$.
\smallpagebreak Pick $\zeta\in\mathcal{D}$ to the left of $\gamma$.
As the vertical curve $\gamma_2\subset\mathcal{D}$ can be taken so
that $\zeta\in R$, we see that $f$ has standard behavior in the whole
part of $\mathcal{D}$ situated to the left of $\gamma$.
\subsection{Behavior of $f$ to the right of $\sigma_1$}
\label{ex:Sl}
First, by means of the Stokes Lemma, Lemma~\ref{st-lm}, we show that
$f$ has standard behavior to the right of $\sigma_1$ in its small
neighborhood.
\smallpagebreak Let $\tilde V_1$ be a sufficiently small constant
neighborhood of $\sigma_1$ and set $V_1=\tilde V_1\cap S_Y$. Show that
$f$ has standard behavior in $V_1\setminus \sigma_1$. The Stokes lines
$\sigma_1$, $\overline{\sigma_1}$ and $[\zeta_{2n-1}^-, \zeta_{2n}^-]$
divide $V_1$ into three sectors.  By the previous steps, we know that
$f$ has the standard behavior in $V_1$ outside the sector
$\mathcal{S}$ bounded by the Stokes lines $\sigma_1$ and
$[\zeta_{2n-1}^-,\zeta_{2n}^-]$. The Stokes line $\sigma_1$ is
vertical. By Lemma~\ref{im-kappa:sign}, in $V_1$, to the left of
$\sigma_1$, the imaginary part of the complex momentum is positive;
thus, the Stokes Lemma implies that $f$ has standard behavior inside
$V_1\setminus\sigma_1$. Recall that the leading term of the
asymptotics of $f$ in $\mathcal{S}$ is obtained by analytic
continuation from the rest of $V_1$ around the branch point
$\zeta_{2n-1}^-$ avoiding $\sigma_1$.
\smallpagebreak Having justified the standard behavior of $f$ to the
right $\sigma_1$ in a small neighborhood of $\sigma_1$, we justify it
in in the rest of the subdomain of $\mathcal{D}\cap\{-Y<\im\zeta<0\}$
situated to the right of $\sigma_1$ by means of the Rectangle Lemma.
The argument is similar to that carried out in subsection~\ref{ex:R}.
So, we omit the details.
\subsection{Behavior of $f$ along the interval
$(\zeta_{2n-1}^-,\zeta_{2n}^-)$}
\label{sec:behavior-f-along}
In the previous steps, we have justified the standard behavior of $f$
both above and below the real interval
$(\zeta_{2n-1}^-,\zeta_{2n}^-)$. We show now that $f$ has
standard behavior also in this interval.\\
By section~\ref{ex:Sl}, we know that $f$ has standard behavior in a
neighborhood of $\zeta_{2n-1}^-$ cut along $\sigma_1$.  Moreover, $f$
can not have the standard behavior in a neighborhood $\zeta_{2n}^-$
(as $\zeta_{2n}^-$ is a branch point).  Hence, there exists
$a\in(\zeta_{2n-1}^-,\zeta_{2n}^-]$ such that $f$ has the standard
behavior in a neighborhood of any point in $(\zeta_{2n-1}^-,a)$.
Assume that $a<\zeta_{2n}^-$. Let $\alpha$ be a segment of the line
$\re\zeta=a$ connecting a point $a_1\in\C_-$ to a point $a_2\in\C_+$.
One has $0<\kappa(a)<\pi$. This implies that, if $\alpha$ is small
enough, then, $\alpha$ is a canonical line. The solution $f$ has the
standard behavior to the left of $\alpha$. By the Adjacent Canonical
Domain Principle, $f$ has standard behavior in any local canonical
domain enclosing $\alpha$, thus, in a constant neighborhood of $a$.
So, we obtain a contradiction, and, thus, $a=\zeta_{2n}^-$. This
completes the analysis of $f$ along $(\zeta_{2n-1}^-,\zeta_{2n}^-)$.
\subsection{Behavior of $f$ to the right of $\sigma_2$}
\label{sec:behavior-f-right}
One studies $f$ to the right of $\sigma_2$ in the same way as we have
studied it to the right of $\sigma_1$: first, using the Stokes Lemma,
one justifies the standard behavior to the right of $\sigma_2$, in a
small neighborhood of $\sigma_2$, and, then, applying the Rectangle
Lemma, one proves that $f$ has the standard behavior in the rest of
the subdomain of $\mathcal{D}$ to the right of this neighborhood. We
omit further details.\\
The analysis of $f$ to the right of $\sigma_2$ completes the proof of
the Proposition~\ref{pro:f:global}. \qed
\subsection{Normalization of $f$}
\label{sec:normalization-f}
To fix the normalization of the leading term of the asymptotics of
$f$, we choose the normalization point $\zeta_0$ in $\mathcal{D}$ and,
in a neighborhood of $\zeta_0$, we choose a branch of the function
$\sqrt{k'(\mathcal{E}(\zeta))}$ in the definition of $\Psi_+$.\\
As the normalization point, we take $\zeta_0$ such that
\begin{equation}
  \label{f:norm-point}
  \zeta_{2n-1}^-<\zeta_0< \zeta_{2n}^-
\end{equation}
Inside any spectral band of the periodic operator,
$k'(\mathcal{E})$ does not vanish, and there are no poles of the
Bloch solution $\psi(x,\mathcal{E})$. So,
$\mathcal{E}(\zeta_0)\not\in P\cup Q$, and the solution $\Psi_+$
is well defined.\\
To fix the branch of $\sqrt{k'}$, we note that, inside any spectral
band of the periodic operator, the main branch of the Bloch
quasi-momentum, $k_p$, is real and satisfies $k_p'>0$. So, we fix
$\sqrt{k'}$ so that
\begin{equation}
  \label{f:q:choice}
  \sqrt{k'(\mathcal{E}(\zeta))}>0,\quad
  \zeta_{2n-1}^-<\zeta<\zeta_{2n}^-.
\end{equation}


%
\section{The consistent basis}
\label{basis}
Up to now, we have constructed $f$, one consistent solution
to~\eqref{G.2} with known asymptotic behavior in the domain
$\mathcal{D}$. We now construct another consistent solution $f^*$ so
that $(f,f^*)$ form a consistent basis.
\subsection{Preliminaries}
\label{sec:preliminaries-1}
For $\zeta\in{\mathcal D}^*$, the symmetric to $\mathcal D$ with
respect to the real line, we define
\begin{equation}
  \label{f-ast}
  f^*(x,\zeta,E)=\overline{f(x,\overline{\zeta},\overline{E})}.
\end{equation}
As $W$ is real analytic, the function $f^\ast$ is also a solution
of~\eqref{G.2}; it satisfies the consistency condition as $f$ does. In
the next subsections, we first study its asymptotic behavior; then, we
compute the Wronskian $w(f,f^*)$. We show that, in $S_Y$, it has the
form $\Const\,(1+o(1))$. Here, $\Const$ is a non-zero constant, and
$o(1)$ is a function which can depend on $\zeta$. Finally, we modify
the solution $f$ so that it still have the standard behavior in
$\mathcal D$, and $w(f,f^*)$ be constant.
\subsection{The asymptotics of $f^*$}
Note that $\mathcal{D}\cap \mathcal{D}^*$ contains the interval
$\mathfrak{z}=(\zeta_{2n-1}^-,\zeta_{2n}^-)\subset\R$. One has
\begin{Le}
  \label{le:f*}
  In $\mathcal{D}^*$, the solution $f^*$ has the standard behavior
  \begin{equation}
    \label{f*:as}
    f\sim e^{-\frac i\varepsilon\int_{\zeta_0}^\zeta\kappa_*
      d\zeta}\,\Psi_{-,*}(x,\zeta,\zeta_0),
  \end{equation}
  where
  \begin{itemize}
  \item $\kappa_*$ is the branch of the complex momentum which
    coincides with $\kappa$ on $\mathfrak{z}$ and is analytic in
    $\mathcal{D}^*$,
  \item $\Psi_{-,*}$ is the canonical Bloch solution which coincides
    with $\Psi_-$ (corresponding to $\Psi_+$ from the asymptotics of
    $f$) on $\mathfrak z$ and is analytic in $\mathcal{D}^*$.
  \end{itemize}
\end{Le}
\demo Recall that, by Proposition~\ref{pro:2}, $f$ has the standard
behavior~\eqref{f:as} in the domain $\mathcal D$.\\
The statement of Lemma~\ref{le:f*} follows from
Proposition~\ref{pro:2}, the definition of $f^*$ and the relation
\begin{equation}
  \label{basis:eq:1}
  \overline{\exp\left(\frac i\varepsilon
      \int_{\zeta_0}^{\bar\zeta}\kappa d\zeta\right)\,
    \Psi_+(x,\bar\zeta,\zeta_0)}=
  \exp\left(-\frac i\varepsilon
    \int_{\zeta_0}^{\zeta}\kappa_*
    d\zeta\right)\,\Psi_{-,*}(x,\zeta,\zeta_0).
  \quad\quad \zeta\in \mathcal{D}^*.
\end{equation}
Let us prove this relation. As both the right and left hand sides
of~\eqref{basis:eq:1} are analytic in $\zeta$, it suffices to
check~\eqref{basis:eq:1} along the interval $\mathfrak z$. Recall that
the interval $[\zeta_{2n-1}^-,\zeta_{2n}^-]$ is a connected component
of $Z$. This implies that
\begin{equation}
  \label{basis:eq:2}
  \overline{\kappa(\zeta)}=\kappa(\zeta),\quad
  \overline{\psi_+(x,\mathcal{E}(\zeta))}=
  \psi_-(x,\mathcal{E}(\zeta)),\quad
  \zeta\in \mathfrak z,
\end{equation}
where $\psi_\pm(x,\mathcal{E})$ are two different branches of the
Bloch solution $\psi(x,\mathcal{E})$.\\
As $\zeta_0$ satisfies~\eqref{f:norm-point},
relation~\eqref{basis:eq:1} follows from the first relation
in~\eqref{basis:eq:2} and the relation
\begin{equation}
  \label{basis:eq:3}
  \overline{\Psi_+(x,\zeta,\zeta_0)}=\Psi_-(x,\zeta,\zeta_0),\quad
  \zeta\in\mathfrak z.
\end{equation}
To check~\eqref{basis:eq:3}, we recall that $\Psi_\pm$ are defined in
section~\ref{CBS} by formula~\eqref{canonicalBS}. Therefore,
relation~\eqref{basis:eq:3} follows from~\eqref{f:q:choice}, the
second relation in~\eqref{basis:eq:2} and the last property of
$\omega$ listed in the section~\ref{sec:Omega}. This completes the
proof of Lemma~\ref{le:f*}. \qed
\subsection{The Wronskian of $f$ and $f^*$}
\label{sec:wronskian-f-f}
The solutions $f$ and $f^*$ are analytic in the strip $S_Y$. Here, we
study their Wronskian. As both $f$ and $f^*$ satisfy
condition~\eqref{consistency:1}, the Wronskian is $\varepsilon$-periodic
in $\zeta$. One has
\begin{Le}
  \label{basis:Wronskian}
  The Wronskian of $f$ and $f^*$ admits the asymptotic representation:
  \begin{equation}
    \label{basis:Wronskian:as}
    w(f,f^*)= w(\Psi_+,\Psi_-)|_{\zeta=\zeta_0}+g,\quad \quad \zeta\in
    S_Y.
  \end{equation}
  Here, $g$ is a function analytic in $S_Y$, such that, along the real
  line, $\re g=0$. Moreover, $g=o(1)$ locally uniformly in any compact
  of $S_Y$ provided that $E$ is in a sufficiently small complex
  neighborhood of $E_0$.
\end{Le}
\begin{Rem} 
  \label{rem:1}
  Note that
  \begin{enumerate}
  \item $w(\Psi_+,\Psi_-)|_{\zeta=\zeta_0}\ne 0$ (as
    $\mathcal{E}(\zeta_0)\not\in P\cup Q$, see~\eqref{f:norm-point}
    and the comments to it);
  \item $w(\Psi_+,\Psi_-)|_{\zeta=\zeta_0}\in i\R$ (due
    to~\eqref{basis:eq:3}).
  \end{enumerate}
\end{Rem}
\demo The domain $\mathcal{D}\cap\mathcal{D}^\ast$ contains the
``rectangle'' $R$ bounded by the lines $\sigma_1\cup
\overline{\sigma_1}$, $\sigma_2\cup \overline{\sigma_2}$ and
$\im\zeta=\pm Y$. So, in $R$, the solutions $f$ and $f^*$ have the
standard behavior~\eqref{f:as} and~\eqref{f*:as}. Consider the
functions $\kappa_*$ and $\Psi_{-,*}$ from~\eqref{f*:as}. Their
definitions, see Lemma~\ref{le:f*}, imply that
\begin{equation*}
  \kappa_*=\kappa,\quad \Psi_{-,*}=\Psi_-,\quad\quad \zeta\in R.
\end{equation*}
Using this information and~\eqref{f:as} and~\eqref{f*:as}, one obtains
\begin{equation}
  \label{basis:eq:4}
  w(f,f^*)= w(\Psi_+(\cdot ,\zeta),\Psi_-(\cdot ,\zeta))+g, \quad g=o(1),
  \quad\quad \zeta\in R.
\end{equation}
Being obtained using standard behavior, this estimate is uniform in
$\zeta$ in any compact of $R$ provided $E$ be in a sufficiently small
neighborhood of $E_0$. By~\eqref{Wcanonical}, the first term in the
left hand side of~\eqref{basis:eq:4} coincides with the first term
in~\eqref{basis:Wronskian:as}. So, we only have to check that $g$ has
all the properties announced in Lemma~\ref{basis:Wronskian}. As
$w(f,f^*)$ is $\varepsilon$-periodic, so is $g$. Furthermore, $ig$ is
real analytic as $iw(f,f^*)$ and $i w(\Psi_+,\Psi_-)|_{\zeta=\zeta_0}$
are. This completes the proof of Lemma~\ref{basis:Wronskian}. \qed
\subsection{Modifying $f$}
\label{sec:correcting-f}
As $g$, the error term in~\eqref{basis:Wronskian:as} may depend on
$\zeta$, we redefine the solution $f$:
\begin{equation*}
  f:=f/\nu,\quad
   \nu=\sqrt{1+g/w(\Psi_+,\Psi_-)|_{\zeta=\zeta_0}}.
\end{equation*}
In terms of this new solution $f$, we define the new $f^*$
by~\eqref{f-ast}. These are the basis solutions the monodromy matrix
of which we shall study. For these ``new'' functions $f$ and $f^*$, we
have
\begin{Th}
  \label{thm:basis}
  The solutions $f$ and $f^*$ satisfy the
  condition~\eqref{consistency:1} and
  \begin{equation}
    \label{new-basis:Wronskian}
    w(f,f^\ast)=w(\Psi_+,\Psi_-)|_{\zeta=\zeta_0}.
  \end{equation}
  Moreover, $f$ has the standard behavior~\eqref{f:as} in $\mathcal
  D$, and $f^*$ has the standard behavior~\eqref{f*:as} in
  $\mathcal{D}^*$.
\end{Th}
\demo Let $\zeta$ be in a fixed strip $\{y_1<\im\zeta<y_2\}\subset
S_Y$, and let $\varepsilon$ be sufficiently small. We use
Lemma~\ref{basis:Wronskian} and Remark~\ref{rem:1}. Recall that $g$ is
$\varepsilon$-periodic in $\zeta$. So, $\nu$ is
$\varepsilon$-periodic, and $f$ and $f^*$ remain consistent.
Furthermore, note that $\nu$ is real analytic. This
implies~\eqref{new-basis:Wronskian}. Finally, as $\nu=1+o(1)$, the new
solutions $f$ and $f^*$ still have the ``old'' standard asymptotic
behavior in $\mathcal D$ and $\mathcal{D}^*$ respectively.\qed


%
\section{General properties of the monodromy matrix for the basis
$\{f,f^*\}$}
\label{sec:m-m:general}
\noindent In the previous section (see Theorem~\ref{thm:basis}),
we have constructed $(f,f^*)$, a consistent basis of solutions
of~\eqref{G.2}. If we return to the variables of the initial
equation~\eqref{family}, we get a consistent basis of~\eqref{family}.
The matrix discussed in Theorem~\ref{Th:mat-mon:1} is the monodromy
matrix obtained for this basis. In this short section, we check some
of its properties.
\smallpagebreak Instead of coming back to the initial variables, we
continue to work in the variables~\eqref{new-variables}. The
definition of the monodromy matrix~\eqref{monodromy} takes the form
\begin{equation}
  \label{monodromy:1}
  F(x,\zeta+2\pi)= M(\zeta)F(x,\zeta),\quad
  F=\begin{pmatrix}f(x,\zeta)\\ f^*(x,\zeta)\end{pmatrix},
\end{equation}
and the matrix $M$ becomes $\varepsilon$-periodic in $\zeta$.\\
As the basis solutions $f$ and $f^\ast$ are related by~\eqref{f-ast},
the monodromy matrix has the form~\eqref{Mform}.\\
The definition of the monodromy matrix~\eqref{monodromy:1} implies that
\begin{equation}
  \label{a,b:w}
a(\zeta)\equiv M_{11}(\zeta)=\frac{w(f(x+2\pi,\zeta),f^*(x,
\zeta))}{w(f(x,\zeta),f^*(x, \zeta))},\quad
b(\zeta)\equiv M_{12}(\zeta)=\frac{w(f(x,\zeta),f(x,
\zeta+2\pi))}{w(f(x,\zeta),f^*(x, \zeta))}.
\end{equation}
Finally, we note that the monodromy matrix is analytic in $\zeta$ in
the strip $S_Y$ and in $E$ in a constant neighborhood of $E_0$.
Indeed, as the solutions $f$ and $f^*$ are analytic functions of both
variables, so are the Wronskians in~\eqref{a,b:w}. Moreover,
by~\eqref{new-basis:Wronskian}, the Wronskian in the denominators
in~\eqref{a,b:w} does not vanish. Hence, we have proved the

\begin{Le}
  \label{m-m:general}
  The monodromy matrix corresponding to the basis $\{f,f^*\}$
  satisfies~\eqref{monodromy:1}, is analytic and
  $\varepsilon$-periodic in $\zeta\in S_Y$, analytic in $E$ in a
  constant neighborhood of $E_0$ and has the form~\eqref{Mform}. Its
  coefficients are given by~\eqref{a,b:w}.
\end{Le}
%


%
\section{General asymptotic formulas}
\label{wronskians:gen-as}
\noindent To compute the asymptotics of the monodromy matrix defined
above, we only need to compute the Wronskians in the numerators
in~\eqref{a,b:w}. These Wronskians depend on $\zeta$ and have
different asymptotics in the lower and upper half planes. Rather than
repeating similar computations many times, in the present section, we
obtain a general asymptotic formula for the Wronskian of two solutions
having standard behavior.
\subsection{General setting}
\label{sec:general-setting}
In this subsection, we do not suppose that $W$ be periodic. Fix
$E=E_0$. Assume that $h$ and $g$ are two solutions of~\eqref{G.2}
having the standard asymptotic behavior in regular domains $D_h$ and
$D_g$:
\begin{equation}
  \label{hg:as}
  h\sim e^{\frac i\varepsilon\int_{\zeta_h}^{\zeta}\kappa_h
  d\zeta}\,\Psi_{h}(x,\zeta),\quad
  g\sim e^{\frac i\varepsilon\int_{\zeta_g}^{\zeta}\kappa_g
  d\zeta}\,\Psi_{g}(x,\zeta).
\end{equation}
Here, $\kappa_h$ and $\kappa_g$ are branches of the complex momentum
analytic in $D_h$ and $D_g$, \ $\Psi_h$ and $\Psi_g$ are canonical
Bloch solutions $\Psi_+$ defined on $D_h$ and $D_g$, and $\zeta_h$ and
$\zeta_g$ are the normalization points for $h$ and $g$.\\
As the solutions $h$ and $g$ satisfy the consistency condition, their
Wronskian is $\varepsilon$-periodic in $\zeta$. We now describe the
asymptotics of this Wronskian and of its Fourier coefficients. We
first introduce several simple useful objects.\\
Below, we assume that $D_g\cap D_h$ contains a simply connected
domain, say $d$.
\subsubsection{Arcs}
\label{sec:arcs}
Let $\gamma$ be a regular curve going from $\zeta_g$ to $\zeta_h$ in
the following way: staying in $D_g$, it goes from $\zeta_g$ to some
point in $d$, then, staying in $D_h$, it goes to $\zeta_h$. We say
that $\gamma$ is an arc {\it associated to the triple $h$, $g$ and
  $d$}.\\
As $d$ is simply connected, all the arcs associated to one and the
same triple can naturally be considered as equivalent; we denote them
by $\gamma(h,g,d)$.
\smallpagebreak Let us continue $\kappa_h$ and $\kappa_g$ analytically
along $\gamma(h,g,d)$. The analysis performed in
section~\ref{sec:kappa}, see~\eqref{allbr}, yields, that, for $V$ a
small neighborhood of $\gamma$, one has
\begin{equation}
  \label{kh,kg}
  \kappa_g(\zeta)=\sigma\kappa_h(\zeta)+2\pi m,\quad\quad
  m\in\Z,\quad \sigma\in\{-1,+1\},\quad \zeta\in V.
\end{equation}
We call $\sigma=\sigma(h,g,d)$ {\it the signature} of $\gamma$, and
$m=m(h,g,d)$ {\it the index} of $\gamma(h,g,d)$.\\
\subsubsection{Meeting domains}
\label{sec:meeting-domains}
Let $d$ be as above. We call $d$ a {\it meeting domain}, if, in $d$,
the functions $\im\kappa_h$ and $\im\kappa_g$ do not vanish and are of
opposite sign.\\
Note that, for small values of $\varepsilon$, the increasing and
decreasing of $h$ and $g$ is determined by the exponential factors
$e^{\frac{i}\varepsilon\,\int^\zeta\kappa d\zeta}$. So, roughly, in a
meeting domain, along the lines $\im\zeta=\Const$, the solutions $h$
and $g$ increase in opposite directions.
\subsubsection{The amplitude and the action of an arc}
\label{sec:amplitude-action-an}
We call the integral
\begin{equation*}
   S(h,g,d)=\int_{\gamma(h,g,d)}\kappa_g d\zeta
\end{equation*}
the {\it action} of the arc $\gamma(h,g,d)$. Clearly, the action takes
the same value for equivalent arcs.
\smallpagebreak Assume that $\mathcal{E} (\zeta)\not\in P\cup Q$ along
$\gamma=\gamma(h,g,d)$. Consider the function
$q_g=\sqrt{k'(\mathcal{E}(\zeta))}$ and the $1$-form
$\Omega_g(\mathcal{E}(\zeta))$ in the definition of $\Psi_{g}$.
Continue them analytically along $\gamma$. Put
\begin{equation}
  \label{Aarc}
  A(h,g,\gamma)=\left.\left(q_g/q_h\right)\right|_{\zeta=\zeta_h}\,
  e^{\int_{\zeta_g}^{\zeta_h}\Omega_g},
\end{equation}
We call $A$ {\it the amplitude} of the arc $\gamma$. The first three
properties of $\Omega$ listed in section~\ref{sec:Omega} imply
\begin{Le} 
  \label{le:3}
  The amplitudes of two equivalent arcs $\gamma(h,g,d)$ coincide.
\end{Le}
\subsubsection{Fourier coefficients}
\label{sec:fourier-coefficients-1}
Let $S(d)$ be the smallest strip of the form $\{C_1<\im\zeta<C_2\}$
containing the domain $d$. One has
\begin{Pro}
  \label{pro:w:as}
  Let $d=d(h,g)$ be a meeting domain for $h$ and $g$, and $m=m(h,g,d)$
  be the corresponding index. Then,
  \begin{equation}
    \label{w:as}
    w(h,g)=w_m\,e^{\frac{2\pi im}{\varepsilon}(\zeta-\zeta_h)}(1+o(1))
    ,\quad \zeta\in S(d),
  \end{equation}
  and $w_m$ is the constant given by
  \begin{equation}\label{wm}
    w_m=A(h,g,d)\,e^{\frac{i}\varepsilon\,S(h,g,d)}\,
    w(\Psi_+(\cdot ,\zeta_h),\Psi_-(\cdot ,\zeta_h)),
  \end{equation}
  where $\Psi_+=\Psi_h$ and $\Psi_-$ is ``complementary'' to $\Psi_+$.
  The asymptotics~\eqref{w:as} is uniform in $\zeta$ and $E$ when
  $\zeta$ stays in a fixed compact of $S(d)$ and $E$ in a small enough
  constant neighborhood of $E_0$.
\end{Pro}
\noindent Note that the factor $w_m$ is the leading term of the
asymptotics of the $m$-th Fourier coefficient of $w(h,g)$.
\demo For $\zeta\in D_g$, let $\gamma_g(\zeta)$ be a curve in $D_g$
from $\zeta_g$ to $\zeta$. Similarly, define $\gamma_h(\zeta)$.\\
First, we check that, for $\zeta\in d$, one has
\begin{gather}
  \label{gaf:1}
  e^{\frac
    i\varepsilon\,\int_{\gamma_g(\zeta)}\kappa_g\,d\zeta}=e^{\frac
    i\varepsilon\,S(h,g,d)}\,e^{\frac {2\pi i
      m}\varepsilon\,(\zeta-\zeta_h)}\,\,\left(e^{-\frac
      i\varepsilon\,\int_{\gamma_h(\zeta)}\kappa_h\,d\zeta}\right),\\
  \label{gaf:2} \Psi_g(x,\zeta)= A(h,g,d)\,\Psi_-(x,\zeta),
\end{gather}
where $\Psi_-$ is the canonical Bloch solution ``complementary'' to
$\Psi_+=\Psi_h$ in a neighborhood of $\gamma_h(\zeta)$.\\
As $d$ is a meeting domain, in a neighborhood of $\gamma(h,g,d)$, one
has
\begin{equation}
  \label{gaf:3}
  \kappa_g=-\kappa_h+2\pi\,m
\end{equation}
This implies relation~\eqref{gaf:1}.\\
Check~\eqref{gaf:2}.  Let $\gamma=\gamma(h,g,d)$ be an arc such that
$\mathcal{E}(\gamma)\cap(P\cup Q)=\emptyset$. Continue $q_g$,
$\Omega_g$ and $\psi_g$ analytically along the arc $\gamma(h,g,d)$.
Note that $q_g$ and $q_h$ are two different branches of the function
$\sqrt{k'(E-W(\zeta))}$. So, they differ at most by a constant factor.
Therefore, in a neighborhood of $\zeta_h$, we get
\begin{equation}
  \label{gaf:4}
  \Psi_g(x,\zeta)= A(h,g,d)\, q_h(\mathcal{E}(\zeta))
  e^{\int_{\gamma_h(\zeta)}\Omega_g}\psi_g(x,\mathcal{E}(\zeta)).
\end{equation}
Now, recall that, in a neighborhood of $\zeta_h$, there are only two
branches of $\Omega$ and $\psi$. Denote them by $\psi_\pm$ and
$\Omega_\pm$ so that $\psi_+=\psi_h$ and $\Omega_+=\Omega_h$. Then,
either $\psi_g=\psi_-$ and $\Omega_g=\Omega_-$ or $\psi_g=\psi_+$ and
$\Omega_g=\Omega_+$. To choose between these two variants, we recall
that the Bloch quasi-momentum of a Bloch solution is defined modulo
$2\pi$. Note that $\kappa_h$ is the Bloch quasi-momentum of $\psi_+$,
and $\kappa_g$ is the Bloch quasi-momentum of $\psi_g$.
By~\eqref{gaf:3}, we get $\kappa_g=-\kappa_h\,{\rm mod}\,2\pi$. So,
$\kappa_g$ must be the Bloch quasi-momentum of $\psi_-$. Thus, we have
$\psi_g=\psi_-$ and $\Omega_g=\Omega_-$, and~\eqref{gaf:4} implies
relation~\eqref{gaf:2} in a neighborhood of $\zeta_h$. By analyticity,
it stays valid in $d$.\\
As $d\subset D_h\cap D_g$, both $h$ and $g$ have standard behavior in
$d$. Substituting the asymptotics of $f$ and $g$ into $w(f,g)$, and
using~\eqref{gaf:1} and~\eqref{gaf:2}, one easily obtains
\begin{equation}
  \label{gaf:5}
  w(h,g)=A(h,g,d)\,e^{\frac i\varepsilon\,S(h,g,d)}\,
  e^{\frac {2\pi im}\varepsilon\,(\zeta-\zeta_h)}\,
  (w(\Psi_+(\cdot ,\zeta),\Psi_-(\cdot ,\zeta))+o(1)),\quad\zeta\in d.
\end{equation}
As $w(\Psi_+(\cdot ,\zeta),\Psi_-(\cdot ,\zeta))$ is independent of $\zeta$ and
$\varepsilon$ and is non-zero (see~\eqref{Wcanonical} and comments to
it), we get~\eqref{w:as}. As this asymptotic was obtained using the
standard behavior of $h$ and $g$, it has all the announced uniformity
properties. This completes the proof of Proposition~\ref{pro:w:as}.
\qed
\subsection{The index $m$ and the periods when $W$ is periodic}
\label{sec:case-periodic-w}
Here, we only assume that $W$ is $2\pi$-periodic and real analytic in
$\zeta$ (i.e. we do not assume anything on the critical points of
$W$), and that $E$ is fixed. We describe the computation of the index
$m$ in the special case that one encounters when computing monodromy
matrices.
\subsubsection{Periods}
\label{indices}
Pick $\zeta_0$, a regular point. Consider a regular curve $\gamma$
going from $\zeta_0$ to $\zeta_0+2\pi$. Fix $\kappa$, a branch of the
complex momentum that is continuous on $\gamma$. We call the
couple $(\gamma,\kappa)$ a {\it period}.\\
Let $(\gamma_1,\kappa_1)$ and $(\gamma_2,\kappa_2)$ be two periods.
Assume that one can continuously deform $\gamma_1$ into $\gamma_2$
without intersecting any branching point. This defines an analytic
continuation of $\kappa_1$ to $\gamma_2$. If the analytic continuation
coincides with $\kappa_2$, we say that the
periods are {\it equivalent}.\\
Consider the branch $\kappa$ along the curve $\gamma$ of a period
$(\gamma,\kappa)$. In a neighborhood of $\zeta_0$, the starting point
$\gamma$, one has
\begin{equation}
  \label{sigmam} 
  \kappa(\zeta+2\pi)=\sigma\kappa(\zeta)+2\pi m,\quad
  \sigma\in\{\pm1\},\quad m\in \Z.
\end{equation}
The numbers $\sigma=\sigma(\gamma,\kappa)$ and $m=m(\gamma,\kappa)$
are called the {\it signature} and the {\it index} of the period
$(\gamma,\kappa)$. The numbers $m$ (resp. $\sigma$) coincide for
equivalent periods.\\
Recall that $G$ is the pre-image with respect to $\mathcal{E}$ of the
spectral gaps of $H_0$. One has
\begin{Le}
  \label{G} 
  Let $(\gamma,\kappa)$ be a period such that $\gamma$ begins at a
  point $\zeta_0\not\in G$. Assume that $\gamma$ intersects $G$
  exactly $N$ times ($N\in\N^*$) and that, at all intersection points,
  $W'\ne 0$.  Let $r_1$, $r_2$, \dots, $r_N$ be the values that
  $\re\kappa$ takes consecutively at these intersection points as
  $\zeta$ moves along $\gamma$ from $\zeta_0$ to $\zeta_0+2\pi$. Then,
  \begin{equation}
    \label{sign,m:period}
    \sigma(\gamma,\kappa)=(-1)^N,\quad m(\gamma,\kappa)=
    \frac1{\pi}\,(r_N-r_{N-1}+r_{N-2}-\dots+(-1)^{N-1}r_1).
  \end{equation}
\end{Le}
\demo The image ${\mathcal E} (\gamma)$ of $\gamma$ by
$\mathcal{E}:\,\zeta\mapsto E-W(\zeta)$ is a closed curve that starts
and ends at ${\mathcal E}_0={\mathcal E}(\zeta_0)$. We consider the
curve $\mathcal{E}(\gamma)$ as open at $\mathcal{E}_0$. Along
$\gamma$, we can write $\kappa(\zeta)=k(E-W(\zeta))$ where $k$ is a
fixed analytic branch of the quasi-momentum. So, $\kappa(\zeta_0)$ and
$\kappa(\zeta_0+2\pi)$, the values of the complex momentum at the ends
of $\gamma_0$, are related by the same formula as $k_b$ and $k_e$, the
values of $k$ at the beginning and the end of curve
${\mathcal E}(\gamma_0)$.\\
Since $W'\ne 0$ at the points of intersection of $\gamma_0$ and $G$,
${\mathcal E}(\gamma_0)$ intersects exactly $N$ times spectral gaps
of the periodic operator.\\
As the values for both $m$ and $\sigma$ coincide for equivalent
periods, it suffices to construct $\zeta_0$ so that
$\im\mathcal{E}_0\neq 0$.\\
Assume that a continuous curve begins at ${\mathcal E}_0$, goes along
a strait line to one of the ends of a gap, then goes around this gap
end along an infinitesimally small circle, and returns back to
${\mathcal E}_0$ along the same strait line. We call such a curve a
simple loop. We distinguish the end and the beginning of the loop
considering it as open at its endpoints. As $\im{\mathcal E}_0\ne 0$,
any simple loop intersects only one gap, namely, the gap around the
end of which it goes.\\
Recall that the ends of the gaps coincide with the branching points of
the Bloch quasi-momentum, and that these branching points are of
square root type. So, in a neighborhood of a branching point, the
corresponding branches of the Bloch quasi-momentum satisfy the
relation
\begin{equation}
  \label{loop}
  k_1(E)+k_2(E)=2 r,
\end{equation}
where $r$ is the common value of these branches at the branching
point. Note that $r$ is equal to the value of the real part of any of
these branches on the spectral gap beginning at the branching point.\\
On a simple loop, fix a continuous branch of the quasi-momentum.
Clearly, formula~\eqref{loop} also relates the values of the
quasi-momentum at the ends of the loop when $r$ is the value of
the quasi-momentum at the branching point inside the loop.\\
Recall that $k$ can be analytically continued onto the whole complex
plane cut along the spectral gaps of $H_0$. Therefore, the value of
$k$ at the end of ${\mathcal E} (\gamma_0)$ is equal to the value of
$k$ at the end of the curve consisting of $N$ simple loops and going
successively around the branch points of $k$ with $k=r_1,r_2,\dots
r_N$. So, taking~\eqref{loop} into account, we get
\begin{equation*}
  k_e=(-1)^N k_b+ 2(r_N-r_{N-1}+r_{N_2}-\dots r_1).
\end{equation*}
This implies~\eqref{sign,m:period} and completes the proof of
Lemma~\ref{G}.\qed
\subsubsection{Indices of periods}
\label{sec:indices-arcs-being}
Let us come back to the computation of the index $m(h,g,d)$. One has
\begin{Le}
  \label{le:m-and-periods} 
  Let $\gamma=\gamma(h,g,d)$ be an arc such that
  $\zeta_h=\zeta_g+2\pi$. If, in a neighborhood of $\zeta_g$,
  \begin{equation}
    \label{kappa:h,g,period}
    \kappa_g(\zeta)=s\cdot \kappa_h(\zeta+2\pi),
  \end{equation}
  where $s$ is either ``+'' or ``-'', then,
  \begin{equation}
    \label{arcs-periods}
    \sigma(h,g,d)=s\cdot \sigma(\gamma,\kappa_g),\quad
    m(h,g,d)=m(\gamma,\kappa_g).
  \end{equation}
\end{Le}
\demo The pair $(\gamma(h,g,d),\kappa_g)$ is a period. So, in a
neighborhood of $\zeta_g$, one has
$\kappa_g(\zeta+2\pi)=\sigma(\gamma,\kappa_g) \kappa_g(\zeta)+2\pi
m(\gamma,\kappa_g)$. This and~\eqref{kappa:h,g,period} imply that
$\kappa_g(\zeta)=s\,\sigma(\gamma,\kappa_g) \kappa_h(\zeta)+2\pi
m(\gamma,\kappa_g)$ in a neighborhood of $\zeta_h$. This implies the
relations~\eqref{arcs-periods}. \qed


%
\section{Asymptotics of the monodromy matrix}
\label{M-asymptotics}
We now compute the asymptotics of the coefficients $a$ and $b$ of the
monodromy matrix for the basis $\{f,f^*\}$; in particular, we prove
formulae~\eqref{a,b:up} and~\eqref{a,b:down}. We concentrate on the
case $n$ odd. The computations for $n$ even being similar, we omit
them.\\
Recall that $a$ and $b$ are expressed via the Wronskians by
formulae~\eqref{a,b:w}. We compute these Wronskians (and, thus, $a$
and $b$) using the construction from section~\ref{wronskians:gen-as}.
\subsection{The asymptotics of the coefficient $b$}
\label{b:odd}
By~\eqref{a,b:w}, we have to compute $w(f(\cdot ,\zeta),f(\cdot ,\zeta+2\pi))$.
One applies the constructions of section~\ref{wronskians:gen-as}.
Now, one has
\begin{gather}
  \label{eq:1}
  h(x,\zeta)=f(x,\zeta),\quad g(x,\zeta)=(Tf)(x,\zeta)
  \quad\text{where}\quad(Tf)(x,\zeta)=f(x,\zeta+2\pi);\\
  D_h=\mathcal{D},\quad D_g=\mathcal{D}-2\pi;\\
  \label{zeta0:h,g} 
  \zeta_h=\zeta_0,\quad \zeta_g=\zeta_0-2\pi;\\
  \label{kappa:h,g}
  \kappa_h(\zeta)=\kappa(\zeta),\quad
  \kappa_g(\zeta)=\kappa(\zeta+2\pi).    
\end{gather}
\subsubsection{The asymptotics in the strip $0<\im\zeta<Y$}
\label{sec:asympt-strip-0imz}
\begin{figure}[h]
  \centering
  \includegraphics[bbllx=71,bblly=549,bburx=332,bbury=721,height=5cm]{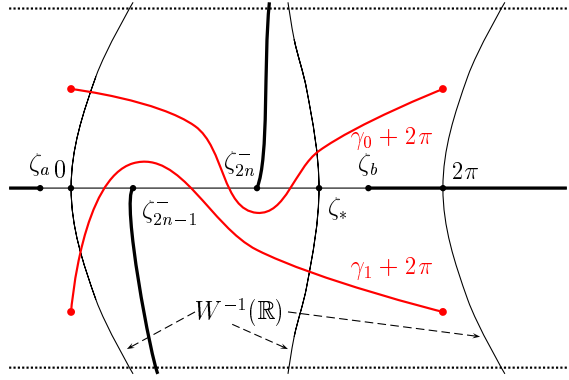}
  \caption{Curves equivalent to $\gamma_0+2\pi$ and $\gamma_1+2\pi$
    (when $m=0$)}
  \label{IBM:fig:2} 
\end{figure}

Let us describe $d_0$, the meeting domain, and
$\gamma_0=\gamma(f,Tf,d_0)$, the arcs used to compute $w(f,Tf)$
in the strip $\{0<\im\zeta<Y\}$.\\
{\it The meeting domain.\/} $d_0$ is the subdomain of the strip
$0<\im\zeta<Y$ between the Stokes lines $\sigma_2-2\pi$ and
$\sigma_2$. Indeed, it follows from Lemma~\ref{im-kappa:sign}
and~\eqref{kappa:h,g} that, in $d_0$, one has $\im\kappa_g=-\im
\kappa_h<0$.\\
{\it The arc.\/} $\gamma_0$ connects the point $\zeta_g$ to $\zeta_h$.
In view of~\eqref{zeta0:h,g}, it defines the period
$(\gamma_0,\kappa_g)$. \\
{\it The index $m(f,Tf,d_0)$.\/} In view of~\eqref{kappa:h,g}, the arc
$\gamma_0$ satisfies the assumption of Lemma~\ref{le:m-and-periods}.
So, $m(f,Tf,d_0)=m(\gamma_0,\kappa_g)$. Due to~\eqref{kappa:h,g},
$m(f,Tf,d_0)=m(\gamma_0+2\pi,\kappa)$. To compute this integer, we use
Lemma~\ref{G}. Therefore, we have to compute $\kappa$ at the
intersections of $\gamma_0+2\pi$ and $G$, the pre-image of the
spectral gaps of $H_0$. The set $G\cap S_Y$ is described in
section~\ref{sec:ZandG} where we have listed all its connected
components.\\
Recall that $m$ takes the same value for all the periods equivalent to
$(\gamma_0+2\pi,\kappa)$. We can deform $\gamma_0+2\pi$ into a curve,
say $\gamma\subset\mathcal{D}$, so that
\begin{itemize}
\item $\gamma$ be to the left of the complex branch of $W^{-1}(\R)$
  starting at $2\pi$ and staying in the upper half-plane,
\item $(\gamma,\kappa)$ be a period equivalent to
  $(\gamma_0+2\pi,\kappa)$,
\item $\gamma$ have the following intersections with the connected
  components of $G$ (for $m=0$, this curve is shown in
  Fig.~\ref{IBM:fig:2}): it once intersects the complex branch of
  $W^{-1}(\R)$ going upwards from $0$, once the interval
  $(\zeta_{2n}^-,\zeta_b)$ (the point $\zeta_b$ is defined in
  section~\ref{sec:results-1}), and, once the complex branch of
  $W^{-1}(\R)$ going upwards from $\zeta^*$.
\end{itemize}
%
%
%
Recall that $\re\kappa(\zeta)$ is constant on any connected component
of $G$. Therefore,
\begin{equation*}
  \begin{split}
    m(\gamma_0+2\pi,\kappa)=m(\gamma,\kappa)
    &=\frac1\pi\left(\left.\re\kappa(\zeta)\right|_{\zeta_{2n-1}^-}-
      \left.\re\kappa(\zeta)\right|_{\zeta_{2n}^-}+
      \left.\re\kappa(\zeta)\right|_{\zeta\in
        \mathfrak{g}_{n+m}+i0}\right)\\
    &=\frac1\pi\left(0-\pi+\left.\re\kappa(\zeta)\right|_{\zeta\in
        \mathfrak{g}_{n+m}+i0}\right).
  \end{split}
\end{equation*}
To compute the last term in this formula, we recall that, in $D_{p,\rm
  left}$, the part of the domain $D_p$ (see
section~\ref{sec:complex-momentum-1}) situated to the left of the
Stokes line $\sigma_2$, one has $\kappa=\kappa_p-\pi (n-1)$. In the
domain $D_{p,{\rm right}}$, the part of $D_p$ situated to the right of
$\sigma_2$, $\kappa$ is obtained by analytic continuation from
$D_{p,{\rm left}}$ around the branch point $\zeta_{2n}^-$ passing
below this branch point. As $\kappa_p(\zeta_{2n}^-)=\pi n$, for $\zeta
\in D_{p,{\rm right}}$, one has $\kappa(\zeta)=(2\pi n-\kappa_p
(\zeta))-\pi(n-1)=\pi(n+1)-\kappa_p(\zeta)$. Along
$\mathfrak{g}_{n+m}$, one has $\re\kappa_p=\pi(n+m)$; hence, we get
\begin{equation}
  \label{m:0:odd}
  m(\gamma_0+2\pi,\kappa)=\frac1\pi(0-\pi+[\pi(n+1)-\pi(n+m)])=- m.
\end{equation}
{\it The result.\/} Now, Proposition~\ref{pro:w:as},
formula~\eqref{a,b:w} for $b$ and formula~\eqref{new-basis:Wronskian}
imply formula~\eqref{a,b:up} for $b$ with
\begin{equation}
  \label{F:b:+:odd}
  b_{-m}=A(f,Tf,d_0)\, e^{\frac {i}\varepsilon
  S(f,Tf,d_0)+\frac{2\pi i m\zeta_0}\varepsilon},\quad
  (Tf)(x,\zeta)=f(x,\zeta+2\pi).
\end{equation}
\subsubsection{Asymptotics of  $b$ below the real line}
\label{b:below:odd}
Describe $d_1$, the meeting domain, and compute the index
$m(f,Tf,d_1)$ needed to get the asymptotics of $w(f,Tf)$ in the
strip $\{-Y<\im\zeta<0\}$.\\
{\it The meeting domain.\/} $d_1$ is the subdomain of the strip
$-Y<\im\zeta<0$ situated between the Stokes lines $\sigma_1-2\pi$
and $\sigma_1$.\\
{\it The index.\/} The arc $\gamma_1=\gamma(f,Tf,d_1)$ again defines a
period, and $m(f,Tf,d_1)=m(\gamma_1+2\pi,\kappa)$. The curve defining
a period equivalent to $(\gamma_1+2\pi,\kappa)$ is shown in
Fig.~\ref{IBM:fig:2}. As in the sequel of this computation we only use
this curve, we call it $\gamma_1+2\pi$. To compute the index of this
period, we compute $\re\kappa$ at the intersections $\gamma_1+2\pi$
and $G$.\\
We can assume that $\gamma_1+2\pi$ satisfies:
\begin{itemize}
\item it is situated to the left of the complex branch of $W^{-1}(\R)$
  in $\C_-$ starting at $2\pi$,
\item it has the following intersections with $G$: it once intersects
  the complex branch of $W^{-1}(\R)$ going downward from $0$, once the
  interval $\mathfrak{g}_{n-1}$ and once the complex branch of
  $W^{-1}(\R)$ going downward from $\zeta^*$.
\end{itemize}
We get
\begin{equation*}
  m(\gamma_1+2\pi,\kappa)=
  \frac1\pi\left.\re\kappa(\zeta)
  \right|_{\zeta\in \mathfrak{g}_{n+m}-i0}.
\end{equation*}
We have used the fact that the interval $\mathfrak{g}_{n-1}$ and the
complex branch of $W^{-1}(\R)$ going downward from $0$ belong to the
same connected component of $W^{-1}(\R)$. To finish the computation,
we introduce the domain $D_p^*$, the symmetric of $D_p$ with respect
to the real line. In $D_{p,\rm right}^*$, the part of this domain
situated to the right of $\sigma_1$, $\kappa$ can be viewed as the
analytic continuation of $\kappa_p-\pi(n-1)$ from $D_p$ across the
interval $\mathfrak{z}_n^-$.  Along the interval $\mathfrak{z}_{n}^-$,
$\kappa_p$ is real. So, for $\zeta\in D_{p,\rm right}^*$,
$\kappa(\zeta)=\overline{\kappa_p(\overline{\zeta})} -\pi(n-1)$, and
\begin{equation}
  \label{m:1:odd}
  m(\gamma_1+2\pi,\kappa)=\frac1\pi\left(
    \left.\re\kappa_p(\zeta)\right|_{\zeta\in
      \mathfrak{g}_{n+m}+i0}-\pi(n-1)\right)=
  \frac1\pi(\pi(n+m)-\pi(n-1))=m+1
\end{equation}
{\it The result.\/} Now, Proposition~\ref{pro:w:as},
formula~\eqref{a,b:w} for $b$ and~\eqref{new-basis:Wronskian} imply
formula~\eqref{a,b:down} for $b$ with
\begin{equation}
  \label{F:b:-:odd}
  b_{m+1}=A(f,Tf,d_1)\, e^{\frac {i}\varepsilon
  S(f,Tf,d_1)-\frac{2\pi i (m+1)\zeta_0}\varepsilon},\quad
  (Tf)(x,\zeta)=f(x,\zeta+2\pi).
\end{equation}
%
%
\begin{figure}[h]
  \centering
  \includegraphics[bbllx=71,bblly=549,bburx=389,bbury=721,height=5cm]{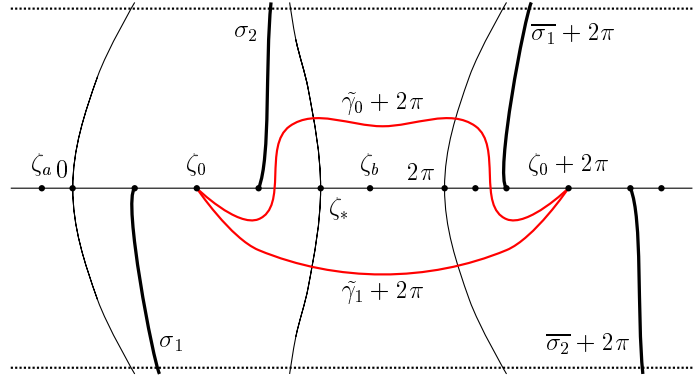}
  \caption{ Curves $\tilde \gamma_0+2\pi$ and $\tilde\gamma_1+2\pi$
    (when $m=0$)}
  \label{a:arcs}
\end{figure}
%
\subsection{The asymptotics of the coefficient $a$}
\label{a:odd}
The computations of the coefficient $a$ following the same scheme as
those of $b$, we only outline them. Now,
\begin{gather}
  h=f^*,\quad g=Tf;\quad\quad
  D_h=\mathcal{D}^*,\quad D_g=\mathcal{D}-2\pi;\\
  \zeta_h=\zeta_0,\quad \zeta_g=\zeta_0-2\pi;\\
  \label{a:kappa:h,g}
  \kappa_h(\zeta)=-\overline\kappa(\bar\zeta),\quad \zeta\in
  D_h,\quad\quad \kappa_g(\zeta)=\kappa(\zeta+2\pi),\quad \zeta\in D_g.
\end{gather}
Recall that the complex momentum is real on $\mathfrak{z}_n^-$, and
that $\zeta_0\in\mathfrak{z}_n^-$. This and
relations~\eqref{a:kappa:h,g} imply that
\begin{equation}
  \label{a:kappas}
  \kappa_g(\zeta)=-\kappa_h(\zeta+2\pi),\quad \zeta\sim\zeta_g.
\end{equation}
\subsubsection{The asymptotics of $a$ above the real line}
\label{sec:asympt-above-real}
In this case, $\tilde d_0$, the meeting domain, is the subdomain of
the strip $\{0<\im\zeta<Y\}$ situated between the lines
$\sigma_2-2\pi$ and $\overline{\sigma_1}$ (which is symmetric to
$\sigma_1$ with respect to $\R$). The arc $\gamma(f^*,Tf,\tilde d_0)$
defines a period $(\tilde\gamma_0,\kappa_g)$; the curve $\tilde
\gamma_0+2\pi$ is shown in Fig.~\ref{a:arcs}. In view
of~\eqref{a:kappas}, one is again in the case of
Lemma~\ref{kappa:h,g,period}, and, by means of Lemma~\ref{G}, one
obtains $m(f^*,Tf,\tilde d_0)=m(\tilde\gamma_0+2\pi,\kappa)=-m$. This
yields formula~\eqref{a,b:up} for $a$ with
\begin{equation}
  \label{F:a:+:odd}
  a_{-m}=A(f^*,Tf,\tilde d_0)\,
  e^{\frac{i}{\varepsilon}\,S(f^*,Tf,\tilde d_0)+\frac{2\pi i
  m\zeta_0}{\varepsilon}}.
\end{equation}
\subsubsection{The asymptotics of $a$ below the real line}
\label{sec:asympt-below-real}
In this case, $\tilde d_1$, the meeting domain, is the subdomain of
the strip $\{-Y<\im\zeta<0\}$ situated between the lines
$\overline{\sigma_2}$ (symmetric to $\sigma_2$ with respect to $\R$)
and $\sigma_1-2\pi$. The arc $\gamma(h,g,\tilde d_1)$ defines a period
$(\tilde \gamma_1, \kappa_g)$; the curve $\tilde \gamma_1+2\pi$ is
shown in Fig.~\ref{a:arcs}. One obtains $m(f^*,Tf,\tilde d_1)=m+1$.
This yields formula~\eqref{a,b:down} for $a$ with
\begin{equation}
  \label{F:a:-:odd}
  a_{m+1}=A(f^*,Tf,\tilde d_1)\,
  e^{\frac i\varepsilon\,S(f^*,Tf,\tilde d_1)-\frac{2\pi i (m+1)\zeta_0}\varepsilon}.
\end{equation}


%
\section{Iso-energy curve}
\label{sec:curves}
\noindent The iso-energy curve $\Gamma$ is defined
by~\eqref{isoen}. In this formula, $\mathcal{E}(\cdot )$ is the dispersion
law for the periodic operator~\eqref{Ho} i.e. the function inverse to
the Bloch quasi-momentum ($\mathcal{E}=\mathcal{E}(k)$ if and only if
$k$ is the value of one of the branches of $k$ when the spectral
parameter is equal to $\mathcal{E}$).  We begin with a simple general
observation:
\begin{Le}
  \label{Gamma:sym}
  The iso-energy curve $\Gamma$ is $2\pi$-periodic in $\zeta$- and
  $\kappa$-directions; it is symmetric with respect to any of the
  lines $\kappa=\pi m$, \ $m\in \Z$.
\end{Le}
\demo The periodicity in $\zeta$ follows from the one of $W$. Fix
$k_0\in\C$. The list~\eqref{eq:55} shows that $\mathcal{E}(k)$ takes
the same value for all $k=\sigma k_0+2\pi m$, where $\sigma\in\{\pm
1\}$, and $m\in\Z$. This implies the periodicity and the symmetries in
$\kappa$.\qed
\smallpagebreak Now, for $W$ satisfying (H) and for $E$ in $J$, an
interval satisfying (A1) -- (A3), we discuss the iso-energy
curves~\eqref{isoenr} and~\eqref{isoen} and obtain the
estimates~\eqref{Fm}.
\subsection{Real iso-energy curve: the proof of Lemma~\ref{realbranches}}
\label{GammaR}
A point $(\zeta,\kappa)\in\R^2$ belongs to $\Gamma_\R$ if and only if
$\kappa$ is the value of one of the branches of the complex momentum
at $\zeta$. Recall that the intervals $\mathfrak{z}\in\mathcal{Z}$ are
pre-images (with respect to $\mathcal E$) of spectral bands. The
complement of these intervals in $(0,2\pi)$ is mapped by $\mathcal E$
into spectral gaps.  So, on $(0,2\pi)$, $\kappa(\zeta)$ takes real
values only on the intervals of ${\mathcal Z}$. Therefore, in the
strip $\{0\le \kappa\le 2\pi\} $, the connected components of
$\Gamma_\R$ are situated above the intervals $\mathfrak{z}\in
\mathcal{Z}$ (``above'' refers to
the projection $\Pi:\,(\zeta,\kappa)\in\R^2\to\zeta\in\R$). \\
Pick $j\in\{n,n+1\dots n+m\}$, and $\sigma\in\{\pm\}$. Consider the
part of $\Gamma_\R$ above the interval
$\mathfrak{z}:=\mathfrak{z}_j^\sigma$.\\
Recall that $\mathcal E$ bijectively maps $\mathfrak{z}$ onto the
$j$-th spectral band of $H_0$. So, there exists $\kappa_0$, a branch
of the complex momentum, continuous on the interval $\mathfrak{z}$,
and mapping it monotonously onto the interval $[\pi(j-1),\pi j]$ so
that $\kappa_0(\zeta_{2j-1}^\sigma)=\pi(j-1)$
and $\kappa_0(\zeta_{2j}^\sigma)=\pi j$.\\
On the interval $[\pi(j-1),\pi j]$, let $Z_{\mathfrak z}$ be the
inverse of $\zeta\mapsto\kappa_0(\zeta)$. We continue $\kappa\mapsto
Z_{\mathfrak{z}}$ to the real line so that it be $2\pi$ periodic and
even. Recall that all the values of all the branches of the complex
momentum at $\zeta\in\mathfrak{z}$ are given by the
list~\eqref{allbr}. This and the definition of $Z_{\mathfrak z}$ imply
that the points of $\Gamma_\R$ above $\mathfrak{z}$ are points of the
graph of $Z_{\mathfrak{z}}$ and reciprocally.\\
All the properties of the function $Z_{\mathfrak z}$ announced in
Lemma~\ref{realbranches} follow directly from this construction.
\smallpagebreak To prove that the connected components of $\Gamma_\R$
depend continuously on $E$, it suffices to check that each of the
functions $Z_{\mathfrak z}$ depends continuously on $E\in J$. Pick
$\mathfrak{z}\in\mathcal Z$. As $W'(\zeta)\ne 0$ for all $\zeta\in
\mathfrak{z}$, the continuity of $E\mapsto Z_{\mathfrak z}$
immediately follows from the Local Inversion Theorem and the
definition of the iso-energy curve~\eqref{isoenr}. This completes the
proof of Lemma~\ref{realbranches}. \qed
\subsection{Loops on the complex iso-energy curve}
\label{s:cl-cur}
Here, we discuss closed curves in $\Gamma$.
\subsubsection{An observation}
\label{sec:an-observation}
We define the intervals $\mathfrak{g}\in\mathcal{G}$ as in
section~\ref{res:cl-cur}. We shall use
\begin{Le}
  \label{cl-cur}
  Pick $\mathfrak{g}\in{\mathcal G}$. Let $V(\mathfrak{g})$ be complex
  neighborhood of $\mathfrak{g}$ sufficiently small so that it
  contains only two branch points of $\kappa$, namely, the ends of
  $\mathfrak{g}$.  Let $\kappa$ be a branch of the complex momentum
  analytic in a sufficiently small neighborhood of a point of
  $V(\mathfrak{g}) \setminus\mathfrak{g}$. Then, $\kappa$ can be
  analytically continued to the domain $V(\mathfrak{g})\setminus
  \mathfrak{g}$ to a single valued function. The analytic continuation
  satisfies
  \begin{equation}
    \label{kVg}
    \kappa(\overline{\zeta})=\overline{\kappa(\zeta)},\quad \zeta\in
    V(\mathfrak{g})\setminus \mathfrak{g}.
  \end{equation}
\end{Le}
\demo We can continue $\kappa$ to a branch of the complex momentum
analytic in $V'(\mathfrak{g})$ the simply connected domain obtained
from $V(\mathfrak{g})$ by cutting it, say, along $\R$ from the right
end of $\mathfrak{g}$ to $+\infty$. It suffices to check that the
values of $\kappa$ at the edges of the cut coincide.\\ 
The set $(\R\cap V(\mathfrak{g}))\setminus \mathfrak{g}$ consists of
two intervals. Each of them belongs to $Z$ (the pre-image of the
spectral bands with respect to $\mathcal E$). So, $\kappa$ is real
both on the left of these two intervals and at the edges of the cut.
As $\kappa$ is real on the left interval, one has~\eqref{kVg} in $V'$.
So, the values of $\kappa$ on the edges of the cut satisfy
$\kappa(\zeta+i0)=\overline{\kappa(\zeta-i0)}$, and, therefore, being
real, coincide. This implies Lemma~\ref{cl-cur}. \qed
\subsubsection{The loops}
\label{sec:loops}
Pick $\mathfrak{g}\in{\mathcal G}$. On $V(\mathfrak{g})\setminus
\mathfrak{g}$, fix $\kappa_0$, a single valued analytic branch of the
complex momentum. Consider $G(\mathfrak{g})\subset V(\mathfrak{g})
\setminus\mathfrak{g}$, a curve going once around the interval
$\mathfrak{g}$. One has
\begin{Le} 
  \label{le:4}
  For each $\sigma\in \{\pm 1\}$ and $m\in\Z$, the curve
  \begin{equation*}
    \hat G_{(m,\sigma)}(\mathfrak{g})=\{(\zeta,\kappa):\
    \kappa=\sigma\kappa_0(\zeta)+2\pi m,\ \zeta\in G(\mathfrak{g})\},
    \quad m\in\Z,
  \end{equation*}
  is a closed curve on $\Gamma$. It connects the two connected
  components of $\Gamma_\R$ that project onto the intervals of $Z\cap
  \R$ adjacent to $\mathfrak{g}$.
\end{Le}
\demo As $\kappa_0$ is univalent on $G(\mathfrak{g})$, the curve $\hat
G_{(0,0)}$ is a closed curve on $\Gamma$. This and
Lemma~\ref{Gamma:sym} imply that all the curves $\hat G_{(\sigma,m)}$
are loops in $\Gamma$. As $G(\mathfrak{g})$ intersects the intervals
of $Z\cap \R$ adjacent to $\mathfrak{g}$, $\hat G$ connects the two
connected components of $\Gamma_\R$ that project onto these
intervals.\qed
\subsection{Tunneling coefficients}
\label{t-coeff}
\smallpagebreak Pick $\mathfrak{g}\in\mathcal G$. Fix an analytic
branch $\kappa$ of the complex momentum on $V(\mathfrak{g})$. Define
the action $S(\mathfrak{g})=i\oint_{ G(\mathfrak{g})}\kappa d\zeta$.
To study its properties, we use
\begin{Le}
  \label{le:S(g)} 
  Let $E\in J$. If $G(\mathfrak{g})$ is positively oriented, then
  \begin{equation}
    \label{oint-int}
    S(\mathfrak{g})=\pm 2\,\int_{\mathfrak{g}\pm i0} \im \kappa d\zeta,
  \end{equation}
  where, in the left hand side, one integrates in the increasing
  direction on the real axis.
\end{Le}
\demo Deform the integration contour $G(\mathfrak{g})$ so that it go
around $\mathfrak{g}$ just along it. Then, relation~\eqref{oint-int}
follows directly from~\eqref{kVg}. \qed
\smallpagebreak This lemma immediately implies
\begin{Cor}
  \label{cor:2}
  Let $E\in J$. Then, 
  \begin{enumerate}
  \item $S(\mathfrak{g})$ is real and non-zero;
  \item as a functional of the branch $\kappa$, it takes only two
    values that are of opposite sign.
  \end{enumerate}
\end{Cor}
\demo Inside any spectral gap, the imaginary part of no branch of the
Bloch quasi-momentum vanishes. Hence, the first statement follows
from~\eqref{oint-int}. The second one follows from~\eqref{allbr}
listing all the branches continuous on the integration contour.\qed
\smallpagebreak In the sequel, we choose the branch $\kappa$ so that,
on $J$, \ $S(\mathfrak{g})$ be positive. $S(\mathfrak{g})$ is called
{\it the tunneling action}.
\subsection{Obtaining estimates~\eqref{Fm}}
\label{sec:obta-estim-eqreffm}
All the estimates in~\eqref{Fm} are obtained in the same way. So, we
prove only the estimate for $b_{-m}$ in the case of $n$ odd. Recall
that we work in $V_0$, a small constant neighborhood of a point
$E_0\in J$.
\smallpagebreak The coefficient $b_{-m}$ is given
by~\eqref{F:b:+:odd}. The definition of the amplitude of an arc,
formula~\eqref{Aarc}, implies that $A(f,Tf,d_1)$ is independent of
$\varepsilon$, continuous in $E$ and does not vanish. So, there are
two positive constants $C_1$ and $C_2$ such that
\begin{equation}
  \label{A:b-m}
  C_1\le |A(f,Tf,d_0)|\le C_2, \quad E\in V_0.
\end{equation}
Let us estimate the factor $\exp\left(\frac{i}\varepsilon S(f,Tf,d_0)
\right)$ for $E\in V_0\cap\R$. Therefore, we choose the arc $\gamma
=\gamma(f,Tf,d_0)$ stretched along the real line and going around the
branch points (between $\zeta_0-2\pi$ and $\zeta_0$, the beginning and
the end of $\gamma$) along infinitesimally small circles. We compute
\begin{equation}
  \label{bm:eq:1}
  \left|\exp\left(\frac {i}\varepsilon S(f,Tf,d_0)\right)\right|=
  \exp\left(-\frac1\varepsilon
    \int_{\gamma}\im\kappa_gd\zeta\right)=
  \exp\left(-\frac1\varepsilon \sum_{\mathfrak{g}\in
      \tilde{\mathcal{G}}} \int_{\mathfrak{g}}\im\kappa_g d\zeta\right),
\end{equation}
where $\tilde{\mathcal{G}}$ consists of all the connected components
of $G\cap\R$ between $\zeta_0-2\pi$ and $\zeta_0$, i.e. of all the
intervals $\mathfrak{g}_j^\pm-2\pi$, the interval $g_{n+m}-2\pi$ and
the interval $g_{n-1}$.
\smallpagebreak Using~\eqref{kappa:h,g}, one easily checks that, in
the right hand side of~\eqref{bm:eq:1}, $\im \kappa_g<0$ inside each
of the intervals of integration (which are segments of the arc
$\gamma$).
\smallpagebreak Due to the periodicity of $W$, we can write
\begin{equation}
  \label{bm:eq:2}
  \left|\exp\left(\frac {i}\varepsilon S(f,Tf,d_0)\right)\right|=
  \exp\left(-\frac1\varepsilon \sum_{\mathfrak{g}\in \mathcal{G}}
    \int_{\mathfrak{g}}\im\kappa d\zeta\right),
\end{equation}
where, on each interval of integration, $\kappa$ is any continuous
branch of the complex momentum such that $\im\kappa<0$. By means of
Lemma~\ref{le:S(g)}, we check that, up to the sign, the expression
$-2\int_{\mathfrak{g}}\im\kappa d\zeta$ is equal to $S(\mathfrak{g})$,
the tunneling action. As both are positive, they coincide. Therefore,
$\D\left|\exp\left(\frac{i}\varepsilon S(f,Tf,d_0)
  \right)\right|=\prod_{\mathfrak{g}\in
  \mathcal{G}}(t(\mathfrak{g}))^{-1}$.  This and~\eqref{A:b-m} imply
the estimate for $b_{-m}$ announced in~\eqref{Fm}.


%
\def\cprime{$'$}

\end{document}